\documentclass[preprint,showpacs]{revtex4}

\usepackage{graphicx}
\usepackage{latexsym}

\def\bx{\mbox{\boldmath $x$}}
\def\bX{\mbox{\boldmath $X$}} 

\def\bk{\mbox{\boldmath $k$}}
\newcommand\ba{\begin{array}}
\newcommand\ea{\end{array}}
\newcommand\ben{\begin{equation}}
\newcommand\een{\end{equation}}
\newcommand\bea{\begin{eqnarray}}
\newcommand\eea{\end{eqnarray}}
\newcommand{\al}{\alpha}  
\newcommand{\be}{\beta} 
\newcommand{\ga}{\gamma}
\newcommand{\de}{\delta}
\newcommand{\ep}{\epsilon}

\newcommand{\ka}{\kappa}
\newcommand{\la}{\lambda}
\newcommand{\rh}{\rho}
\newcommand{\si}{\sigma}
\newcommand{\ta}{\tau}

\newcommand{\Ga}{\Gamma}
\newcommand{\De}{\Delta}

\newcommand{\La}{\Lambda}

\newcommand{\Om}{\Omega}

\newcommand{\half}{\frac{1}{2}}
\newcommand{\A}{{\mathcal{A}}}
\newcommand{\C}{{\mathcal{C}}}
\newcommand{\D}{{\mathcal{D}}}
\renewcommand{\L}{{\mathcal{L}}}

\newcommand{\vev}[1]{\langle#1\rangle}
\newcommand{\pa}{\partial}

\newcommand{\detbar}{\overline{\det}}

\begin{document}

\date{July 2002, Revised December 2002}

\preprint{SUSX-TH/02-013, arXiv:hep-ph/0207267}

\title{Level Set Method for the Evolution of Defect and Brane Networks}
\author{Mark Hindmarsh}
\email{m.b.hindmarsh@sussex.ac.uk}
\affiliation{
Centre for Theoretical Physics, University of Sussex, 
Brighton BN1 9QJ, U.K.
}%

\begin{abstract}  

A theory for studying the dynamic scaling properties of branes and relativistic
topological defect networks is presented.  The theory, based on a relativistic
version of the level set method, 
well-known in other contexts, possesses
self-similar ``scaling'' solutions, for which one can calculate many quantities
of interest.  Here, the length and area densities of cosmic strings and domain
walls are calculated in Minkowski space, and radiation, matter, and
curvature-dominated FRW cosmologies with 2 and 3 space dimensions.  The scaling
exponents agree the naive ones based on dimensional analysis, except for cosmic
strings in 3-dimensional Minkowski space, which are predicted to have a
logarithmic correction to the naive scaling form.  The scaling {\it amplitudes}
of the length and area densities are a factor of approximately 2 lower than
results from numerical simulations of classical field theories.  An expression
for the length density of strings in the condensed matter literature is
corrected.

\end{abstract} 
 
\pacs{98.80.Cq, 11.27.+d, 64.60.Cn}
\maketitle

\section{Introduction} 

The
solutions to some of the most interesting problems in physics depend on a better
understanding of the dynamics of fields far from thermal equilibrium,
particularly in particle cosmology, where we seek mechanisms for generating
a baryon asymmetry \cite{BarAsy}, for generating density fluctuations
\cite{DenFlu}, and perhaps for generating primordial magnetic fields
\cite{MagFie}. Significant advances have been made recently in 
studying non-equilibrium 
dynamics of phase transitions, both in theoretically 
(see e.g.\ \cite{Boyanovsky:1999db} for a review) and numerically, 
where we can now perform real time simulations of a quench 
with leading thermal corrections included \cite{HotAbHiggs,Rajantie:2001ps}.
One aspect is still not yet well 
understood: the approach to equilibrium after a phase transitions of a
field theories with topological defects.

At the same time, the last few years has seen an explosion in theories
involving various kinds of extended objects or branes, both solitonic (like
topological defects in field theory) and fundamental . Most
of the interest has lain in special configurations of branes of various
dimensions, and the spectrum of states in those backgrounds.  However,
an interesting new scenario has emerged in which  
the Universe began with the branes
in thermal equilibrium, the brane gas Universe \cite{BraGas}

Both branes and topological defects in relativistic field theories obey the
same equation of motion (at least for configurations with curvature small
compared to the inverse width or fundamental scale), and so it clear that both may be discussed 
at the same time. Hence the theory presented in this 
paper can be applied to both brane gases and networks of topological defects. 
The general technique is independent of the space-time dimension and the
codimension of the brane, but quantitative predictions must be taken case by
case.  The cases worked out in detail here concerned defects of codimension 1
and 2 in Friedmann-Robertson-Walker (FRW) spacetimes of
dimension $d=3$ and $4$. 

It is believed that when extended topological defects
are formed, self-similar or scaling behaviour emerges at large times, in
which a characteristic length scale of the field configuration, $\xi$,
increases with time as a power law:
$$
\xi(t) \propto t^z.
$$
Dynamic scaling can be seen in the order parameter 
of many condensed matter systems
undergoing rapid quenches, and there are now quite sophisticated
techniques for calculating 
correlation functions of the order parameter \cite{Bra94}.  They fall into two
classes. Firstly, there are those based on a large $N$ expansion, 
where $N$ is the number of components of the order parameter, which are
applicable to Ginzburg-Landau theories.  The second is applicable to
systems with extended topological defects, in which the 
order parameter $\phi$ obeys an equation equation 
$\dot\phi \propto \de F[\phi]/\de\phi$, where $F$ is the Ginzburg-Landau free 
energy.  Allen and Cahn \cite{AllCah79} proposed that the velocity of defects marking a phase boundary  was proportional their local mean
curvature.  This proposal, now termed motion by mean curvature, was later rigorously proved 
\cite{MotMeaCur}.

Relativistic scalar field theories with spontaneously broken global symmetries
(Goldstone models) also exhibit dynamic scaling. Significant progress has
been made on the theory of O($N$) scalar field theories at large $N$, both
classical \cite{Turok:1991qq,FilBra94,Kunz:1996ka} and quantum \cite{Boyanovsky:1998yp}
(at large $N$ the leading order in the quantum theory is 
the same as the classical theory). These works have established a
theoretical basis for the scaling observed in numerical simulations
\cite{Bennett:1993fy,Durrer:1994da,Pen:1994nx}. The theory has also been
used to calculate microwave background and density fluctions. To date,
however, analytic approaches to the dynamics of topological defects are
few.

There are several numerical simulations 
which broadly support the dynamic scaling hypothesis for topological 
defects, including domain walls 
\cite{Press:1989yh,Coulson:1996nv,Larsson:1997sp}, 
gauge strings
\cite{Vincent:1998cx,Moore:2001px}, and global strings
\cite{Ryden:1989vj,Yamaguchi:2000dy}. All the
simulations are consistent with the linear scaling law over the
range of the simulations, although 
Press, Ryden and Spergel suggested that the results
for domain walls would be better fitted by $ \xi \sim t/\ln (t) $; however,
more recent simulations with a larger dynamic range \cite{GarHin02} are not
consistent with the logarithm.

There are also string simulations based on direct 
integration of the equations of motion 
of one-dimensional objects, 
obeying the Nambu-Goto equations, which may be derived as the first
approximation in an expansion in powers of string worldsheet curvature
\cite{Carter:1994ag,Anderson:1997ip,Arodz:1997va}
They do not include any way for energy to be lost from the network, but if one
considers ``infinite''
strings only (strings longer than the causal horizon size),
an approximately  linear scaling law is found \cite{Albrecht:1989mk,StrSimBB,StrSimAS}.
However, the simulations are plagued by 
kinkiness persisting at the resolution of the simulation, associated with the
production of small loops of string,
which does not appear to scale.  It has been suggested that this is because the
natural length scale for loop production is in fact the string width, where
loops would become indistinguishable from large amplitude oscillations in the
field \cite{Vincent:1996rb}.  Indeed, numerical simulations of the fields 
\cite{Vincent:1998cx,Moore:2001px} appear to
support this hypothesis, although the latter authors suggest 
that the ``protoloops''
in their simulation are in fact a transient effect.

A programme to understand analytically the results of the  Nambu-Goto
simulations has developed 
over the years  \cite{Kibble:1984hp,Albrecht:1989mk,KibCop,VelDep1Scale}.
In its simplest form, the model parametrises the string with one length scale
$\xi$, which is defined from the invariant length density of infinite string
$\L_\infty$ through $\L_\infty = 1/\xi^2$. This length density can change in two ways:
through stretching as the strings participate in the Hubble expansion, and
through loop production.  Loop production is parametrised by the so-called
chopping efficiency $c$,  the
fraction of string lost to the network in the timescale $\xi$. The Hubble
stretching depends on the mean square string velocity $v^2$.  The
phenomenological equation is then 
\ben
\dot\xi = H( 1 + v^2) \xi +  c/2 .
\een
Further work \cite{KibCop} introduced two other length scales to describe the
correlation length and the inter-kink distance.  
However, there are many unknown parameters in the model 
which greatly restricts its predictive power, despite attempts to measure them
\cite{Vincent:1996rb}.  A different approach was adopted by Martins and Shellard
\cite{VelDep1Scale} who promoted the r.m.s. string velocity $v$ to a
time-dependent parameter, to model the reduced rate of loop production 
of slower strings.  The velocity-dependent one-scale model equations are 
(neglecting frictional terms) 
\ben
\dot\xi = H( 1 + v^2) \xi +  \tilde{c}v/2, \quad \dot v = -2H v + k(1-v^2)\xi^{-1},
\een  
where $\tilde{c}$ and $k$ are, in the simplest version, constants. 
It is this velocity-dependent one-scale
model which \cite{Moore:2001px} use to make their claim that the production of
loops on the scale of the string width seen in field theory simulations is a
transient.

In this paper a potentially far more powerful analytic technique 
for describing the motion of strings is developed.  The technique was 
outlined in 
\cite{Hindmarsh:1996xv,HinCop97} and 
applied to relativistic domain walls in 2 and 3 space dimensions. 
It is here further extended into a partial treatment of 
$p$-branes in D space dimensions, and fully applied to relativistic
strings in 3 space dimensions.
It is based on the $u$-theory of
Ohta, Jasnow and Kawasaki (OJK) \cite{OhtJasKaw82}, and 
its descendents \cite{DefScale}, which describes the motion of 
defects obeying the Allen-Cahn equation. The relativistic generalisation 
of the Allen-Cahn equation is the Nambu-Goto equation, in which,
loosely speaking, the {\em acceleration} of the defect is proportional
to its local curvature, with proportionality constant $c^2$, where 
$c$ is the speed of light.  More precisely, the Nambu-Goto equation is
equivalent to the requirement that the world-volume of the $p$-brane 
embedded in the $d$-dimensional spacetime has 
zero extrinsic curvature.  How closely defects derived from a field
theory obey this equation is a matter for
debate \cite{Carter:1994ag,Anderson:1997ip,Arodz:1997va,Olum:1999sg}.
 The theoretical approach develops systematic expansions of the geometrical 
equations obeyed by the defect world-volumes in powers of the width
divided by the local curvature, which reduce to the Nambu-Goto equation
in the limit of small curvature.  The approach of Arod\'z \cite{Arodz:1997va} makes it
particularly clear that the Nambu-Goto equation is really a consistency
condition for a smooth defect-like solution to exist.
 
It is therefore plausible that we can forget about the details of the
field theory and concentrate instead on the properties of extremal 
(zero extrinsic curvature) surfaces embedded in higher dimensions. If one 
finds such surfaces, then provided their curvature is small enough one
can be confident that there is a solution of the field equations
representing a smooth defect centred on that surface.
A formalism for studying extremal, and more general, surfaces has been developed over the 
years by Carter \cite{Carter:2000wv}, which makes clear the geometrical 
nature of the Nambu-Goto equations through close attention to the tensorial 
properties.

The present approach 
we introduce scalar fields $u^A$ with the intention that the  
loci of constant $u^A$ should be extremal surfaces: these are the level sets 
of the title. The  
fields can also be interpreted as coordinates normal to the brane surface:
in this sense the approach can be thought of as orthogonal to Carter's.
We derive the equations that the $u^A$ must satisfy, 
which are non-linear, and so therefore do not seem to represent 
an improvement on the original field theory or the Nambu-Goto equations. 
However, one can derive equations for surfaces which are 
{\em on average} extremal, when we average the fields 
with a Gaussian probability distribution. 
With this Gaussian ansatz, one can also calculate analytically
important quantities, such as the brane or defect density.

The results for $(D-1)$-branes (domain walls) are extremely 
encouraging when compared to the numerical simulations 
\cite{Press:1989yh,Coulson:1995nv,Larsson:1996sp}. 
The theory predicts a scaling
law for the area density in 3 dimensions, but not only does it 
predict the scaling exponent, it also predicts the scaling {\em amplitude} to
within a factor of about 2, which is not bad given the
approximations made.  The prediction for $(D-2)$-branes 
in 3 dimensions (strings) is
also challenging: the theory gives a logarithmic scaling violation 
in Minkowski space, with the
length density deoending on conformal time $\eta$ as $\log(\eta)/\eta^2$. Looking for
such scaling violations will be a good way to test the theory, although 
computationally very challenging.

The theory also describes the behaviour of
defects formed from initial conditions with a slight bias 
in the expectation value of the field favouring 
one vacuum over another \cite{Coulson:1995nv,Larsson:1996sp,Casini:2001ai}.  
It is found 
that the defects disappear exponentially fast at a critical conformal 
time $\eta_c$, which scales with the initial bias $U$ as
$\eta_c\sim U^{2/D}$. 
Indeed, part of the motivation for this work was to account for 
this kind of behaviour observed in simulations by Coulson, Lalak and Ovrut 
\cite{Coulson:1995nv} and 
Larsson, Sarkar and White \cite{Larsson:1996sp}.

Finally, in making comparisons with similar results in the condensed matter 
literature, an expression for the length density of 
strings in 3 space dimensions in the condensed matter is corrected 
(see Section \ref{s:AvPAD2}).

In this paper we shall work a 
conformally flat $d$-dimensional Friedmann-Robertson-Walker
space-time with coordinates $x^0,x^1,\ldots,x^D$, such that $d=D+1$. 
The metric is given by  
\ben 
g_{\mu\nu} = a^2(\eta){\rm diag}(-1,\de_{ij}), 
\een 
where $\eta$ is conformal time, giving an affine connection   
\ben 
\Gamma^\rho_{\mu\nu} = (\delta_\mu^\rho\delta_\nu^0  
+ \delta_\nu^\rho\delta_\mu^0 - g_{\mu\nu}g^{\rho 0})(\dot a/a). 
\een

\section{Field equations}

In this section we shall firstly study
model field equations for topological defects
of codimension $N=1$ and $N=2$, which correspond to walls and strings
respectively in $D=3$.  We shall see that we can find approximate 
solutions to the field equations near surfaces of codimension $N$ which 
have zero extrinsic curvature, and whose other curvature radii are 
large compared with the width of the defect.  These results are well known 
and have been shown in various ways in \cite{Carter:1994ag,Anderson:1997ip,Arodz:1997va}, 
but the approach here is slightly 
different and worth exhibiting in some detail for the later sections of the
paper.

\subsection{Domain walls}
Let us first consider a theory with a single scalar field $\phi$, with action
\begin{equation}
S= -\int d^dx\,\sqrt{-g}\left(\half\pa_\mu\phi\pa^\mu\phi + V(\phi)\right),
\end{equation}
from which we derive the field equation
\begin{equation}
-\frac{1}{\sqrt{-g}}\pa_\mu(\sqrt{-g}g^{\mu\nu}\pa_\nu)\phi +
\frac{dV}{d\phi}=0
\end{equation}
We shall suppose that the potential $V$ has the symmetry $\phi\to -\phi$, 
and moreover that its minima are at $\phi=\pm v$, with $V(\pm v) =  0$. 
If we impose the boundary conditions:
\ben
\phi(x^D \to -\infty) = -v, \quad \phi(x^D\to +\infty) = +v,
\een
and make the ansatz
\ben
\pa_\mu\phi(x)=0 \quad (\mu=0,\ldots,D-1), 
\een
then the theory has a one parameter family 
of domain wall solutions, with $\phi=0$ at 
$x^D=X^D$.  If the potential is quartic, 
\ben
V(\phi) = \frac{1}{4}(\phi^2-v^2)^2,
\een
then the solutions are
\ben
\bar\phi(x) = v\tanh [M(x^D - X^D)],
\een
where $M=\sqrt\la v$. Thus the width of the defect is controlled by the 
parameter $M^{-1}$.
The defect can be thought of as centred at $X^D$, where the field vanishes,
with a width parameter $M^{-1}$.

\subsection{Strings}
The simplest theory to exhibit string-like solutions is the Abelian Higgs 
model, which has action
\ben
S = - \int d^dx\sqrt{-g} \left(\frac{1}{4}
F_{\mu\nu}F^{\mu\nu} + D_\mu\phi^*D^\mu\phi + V(\phi)\right),
\een
where $\phi$ is a complex scalar field with covariant derivative
$D_\mu\phi = \pa_\mu\phi - ieA_\mu\phi$.  The potential $V$ is taken to respect
a U(1) symmetry $\phi \to e^{i\al}\phi$, with a circle of minima at 
$|\phi| = v$.  If we impose the boundary conditions in the $\{x^{D-1},
x^D\}$ plane
\ben
\phi(r \to  \infty) = ve^{i\theta}, 
\een
where $r^2 = (x^{D-1})^2 + (x^D)^2$ and $\tan\theta = x^D/x^{D-1}$, then by
continuity 
$\phi$ must vanish somewhere in the plane. If we furthermore 
assume translational invariance in the other $d-2$ directions in spacetime, 
we find a two-parameter family of static string
solutions, labelled by the coordinates
of the centre of the string, $\{X^{D-1},X^{D}\}$.  
In the radial gauge $A_r = 0$ these solutions take the
form
\ben
\bar\phi(x) = f(\rho)e^{i\varphi},
\quad \bar A_i = {1\over e\rho}\hat{\varphi}_i a(m_v\rho),
\quad \bar A_\al = 0,
\een
where $\rho_1 = (x - X)^{D-1}$, $\rho_2 = (x-X)^D$, $\rho^2 = (\rho_1)^2 + (\rho_2)^2$, 
$\tan\varphi = \rho_2/\rho_1$, and $\hat\varphi^i$ is the unit azimuthal vector in 
the $\{x^{D-1}, x^D\}$ plane.
These solutions cannot generally be found analytically, even when the potential
has the renormalisable and gauge invariant form
\ben
V(\phi) = \half \la (|\phi|^2 - v^2/2)^2.
\een
However, they are easily found numerically, and exhibit similar properties to
the domain wall in that away from the centre of the defect the
fields approach their vacuum values exponentially, at rates controlled by the
masses of the fields $m_s = \sqrt{\la}v$ and $m_v = ev$.  Defining a dimensionless 
coordinate $z = m_v\rho$, and $\be = (m_s/m_v)^2 = \la/e^2$, one has \cite{Hindmarsh:1994re}
\ben
f \sim 1 - f_1z^{-1/2}\exp(-\sqrt{\be}z), \quad a \sim 1-a_1z^{1/2}\exp(-z).
\een
In the case $\be > 4$, the asymptotic form of $f$ is $1 - z^{-1}\exp(-2z)$.

Again, the string can be thought of as centred at $\{X^{D-1},X^D\}$, with thickness 
$m_v$, although for light scalars ($\be \ll 1$) there is a thicker scalar core where the 
scalar field asymptotes to its vacuum value.

\subsection{Solutions in curvilinear coordinates}
\label{s:CurvCoord}
These are however rather special solutions with a high degree of 
symmetry.  Let us instead look for (if necessary approximate)
solutions, corresponding 
to defects centred on a more general surface 
$X^\mu(\si^\al)$, with $\al = 0,\ldots,p=D-N$. 
We choose a new set of
coordinates $\xi^\mu = \{\si^\al,u^A\}$, where $A=1,\ldots,N$, with the
intention that the equations of the surfaces can be written
\ben
u^A(x) = 0.
\een
We write the metric in these new coordinates
\ben
G_{\mu\nu} = \left( \ba{cc} \pa_\al x \cdot \pa_\be x  & \pa_\al x \cdot \pa_B x  \cr
		\pa_A x \cdot \pa_\be x	& \pa_Ax \cdot \pa_Bx	\cr  \ea
\right)
= \left( \ba{cc} \ga_{\al\be}  & N_{B\al}   \cr
		N_{A\be}	& G_{AB}	\cr  \ea
\right),
\label{e:Metric}
\een 
where the dot indicates a contraction with respect to the original metric
$g_{\mu\nu}$.  We may choose the coordinates $\xi^\mu$ so that, at 
least at $u^A=0$, the $u^A$ and $\si^\al$ are locally orthogonal, or
\ben
\left.N_{A\be}\right|_{u^A=0} = 0.
\een
In fact, with walls and strings in $D=3$, these are only 3 or 4 conditions 
on the metric respectively, so we know we can make a coordinate transformation
so that this is true everywhere, and not just at $u^A=0$.

Note that the upper left $(p+1)\times(p+1)$ block of $G_{\mu\nu}$, denoted  
$\ga_{\al\be}$ in Eq.\ (\ref{e:Metric}), is the embedding metric on surfaces of constant
$u^A$, which they acquire by virtue of being surfaces embedded in a spacetime with metric
$g_{\mu\nu}$.  

We can also write the inverse metric 
\ben
G^{\mu\nu}
= \left( \ba{cc} \pa \si^\al \cdot \pa \si^\be & \pa \si^\al \cdot \pa u^B  \cr
		\pa u^A \cdot \pa\si^\be       & \pa u^A \cdot \pa u^B	\cr  \ea
\right).
\een 
We define 
\ben
h^{AB} = \pa u^A \cdot \pa u^B,
\een
and use the convention that the indices $\al$, $\be$, etc.\ are raised and
lowered with $\ga_{\al\be}$ and $\ga^{\al\be}$ (defined as the matrix inverse),
and that the indices $A$, $B$, etc.\ are raised and lowered with $h^{AB}$ and
its matrix inverse $h_{AB}$. Hence 
\ben
G_{\mu\nu} = \left( \ba{cc} \ga_{\al\be}  & N_{B\al}  \cr
		N_{A\be}&  h_{AB} + N_{A\be}{N_{B}}^\be \cr  \ea
\right),
\quad
G^{\mu\nu}
= \left( \ba{cc} \ga^{\al\be} + {N_{A}}^\al N^{A\be} & -N^{B\al}  \cr
		-N^{A\be} &  h^{AB} \cr  \ea
\right).
\een
One can show that 
\ben
\det G^{\mu\nu} = \det \ga^{\al\be}\det h^{AB},    
\een
and hence that $G = \ga/ h$, 
where $G=\det G_{\mu\nu}$, $\ga = \det\ga_{\al\be}$, and
$h = \det h^{AB}$.

We have two projectors associated with the constant $u^A$ 
surfaces, one which projects onto the surface and the other which
projects onto the subspace spanned by the vectors $\pa_\mu u^A$.
\ben
P^\mu_{\parallel\nu} = 
\gamma^{\alpha\beta}\partial_\alpha X^\mu
\partial_\beta X_\nu
,
\quad
P^\mu_{\perp\nu} = h_{AB}\pa^\mu u^A \pa_\nu u^B,
\een
with $ P^\mu_{\parallel\nu} + P^\mu_{\perp\nu} = \de^\mu_\nu.$
%
%
%
%
\begin{figure}
\centering
\scalebox{0.5}{\includegraphics{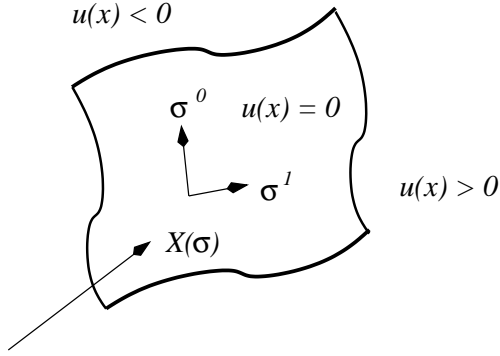}}
\caption{\label{f:geom1} The coordinates $\xi^\mu = \{\si^\al, u^A \}$,
where $\al=0,\ldots,p$ and $A=1,\ldots,N$, which are 
chosen so that $u^A=0$ will be the extremal surface on which 
the topological defect sits. Illustrated is a 1-brane in $2+1$ space-time
dimensions, located at $X^\mu(\si) = x^\mu(\si,0).$}
\end{figure}

Let us study the field equation for the theory of an $N$-component 
scalar field $\Phi$ in these new coordinates: 
\ben
-\frac{1}{\sqrt{-G}}\pa_\al\left( \sqrt{-G} \ga^{\al\be}\pa_\be\right)\Phi 
= \frac{1}{\sqrt{-G}}
\hat\pa_A\left(\sqrt{-G}h^{AB}\hat\pa_B\right)\Phi 
- \frac{dV}{d\Phi},
\label{e:ScaFieEOMNewCoo}
\een
where $\hat\pa_A = \pa_A -{N_A}^\al\pa_\al$
(where we use notation mirroring that of Moss and Shiiki \cite{Moss:1998jf}). 
At the surface $u^A=0$ it 
is possible to make a coordinate transformation amongst the $u^A$
coordinates so that they are orthonormal, that is, $h^{AB} = \de^{AB}$.  

This choice of coordinates is different from the one used in other works
on solving defect equations of motion in curvilinear coordinates 
\cite{Carter:1994ag,Anderson:1997ip,Arodz:1997va}, where coordinates
$\{\si^\al,\rho^A\}$ are constructed away from the surface by setting
\ben
x^\mu(\si^\al,\rho^A) = X^\mu(\si^\al) + \rho^A n_A^\mu(\si^\al),
\label{e:ONcoords}
\een
where ${n_A}^\mu = \left.\pa_Ax^\mu\right|_{u^A=0}$, and 
${n_A}\cdot{n_B} = \de_{AB}$.  The coordinates coincide only when
$N_{A\be} = 0$.
Carter \cite{Carter:2000wv} also uses orthonormal vectors in the surface, 
and is careful to express quantities as space-time tensors.  
Table \ref{t:CarterTranslate} contains a summary which compare his notation and 
conventions with this work.

\begin{table}
\begin{tabular}{c|l|c|c}
\textit{This work} & \textit{Name} & \textit{Carter}  & \textit{Relationship} \\
\hline
${P^\mu_{\parallel\nu}}$  & First fundamental tensor & ${\eta^\mu}_\nu$ & ${P^\mu_{\parallel\nu}} = {\eta^\mu}_\nu$\\ \hline
${P^\mu_{\perp\nu}}$ & Orthogonal projector & ${\perp^\mu}_\nu$ & ${P^\mu_{\perp\nu}} = {\perp^\mu}_\nu$\\ \hline
$\pa_\al x^\mu$ & Tangent vector  & ${\iota_A}^\mu$  & ${\iota_A}^\mu{\iota_B}_\mu = \eta_{AB}$, \\[-1pt]
($\al = 0,\ldots,p$) & & ($ A = 0,\ldots,p$)& $\pa_\al x^\mu\pa_\be x_\mu = \ga_{\al\be}\ne \eta_{\al\be}$ \\\hline
$\pa_\mu u^A$ & Normal vector  & ${\la_X}^\mu$ & ${\la_X}^\mu{\la_Y}_\mu= \de_{XY}$, \\[-1pt]
($A = 1,\ldots,N$)& & ($X = 1,\ldots,N$)&
$\pa_\mu u^A \pa^\mu u^B = h^{AB}\ne\de^{AB}$\\ \hline
$K^A_{\mu\nu}$ & Second fundamental tensor & ${K_{\mu\nu}}^\rho$ & $K^A_{\mu\nu} = {K_{\mu\nu}}^\rho\pa_\rho u^A$\\
\hline
\end{tabular}
\caption{\label{t:CarterTranslate} Comparison of notation and conventions with that 
of Carter \cite{Carter:2000wv}.}
\end{table}

In contrast to previous work, 
it is here more convenient to use the unnormalised $\pa_\mu u^A$ as 
basis vectors, as we are interested in the surfaces 
generated by gaussian random fields
$u^A$, with unconstrained derivatives at $u^A=0$.

We now try to find approximate solutions to Eq.\
(\ref{e:ScaFieEOMNewCoo}).  A promising avenue is to look for solutions 
which are independent of $\si^\al$, in which case Eq.\
(\ref{e:ScaFieEOMNewCoo}) becomes
\ben
\sqrt{h}\pa_A\left(\frac{1}{\sqrt{h}}h^{AB}\pa_B\right)\Phi 
+ K^A\pa_B\Phi - \frac{dV}{d\Phi} = 0.
\een
where  
$K^A$ is the extrinsic curvature of the constant $u^A$ hypersurfaces,
given by
\ben
K^A = \frac{1}{\sqrt{-\ga}}h^{AB}\pa_B(\sqrt{-\ga}).
\een
The ansatz $\Phi=\Phi(u^A)$ can only be self-consistent if both $K^A$ and
$h^{AB}$ are independent of $\si^\al$. This is still a difficult equation to
solve, so the next step is to look  
near surfaces where the extrinsic curvature vanishes.  
Transforming to the orthonormal coordinates (\ref{e:ONcoords}) 
near those surfaces we
have the approximate equations
\ben
- \frac{\pa}{\pa \rho^A}\frac{\pa}{\pa\rho^A}\Phi 
+ \frac{dV}{d\Phi} \simeq 0.
\label{e:DefEqn1}
\een
By ``near'' we mean the region where $|K^A\pa_A\Phi| \ll |\pa_A\pa_A\Phi|$.
Eq.\ \ref{e:DefEqn1} is solved by $\bar\Phi(\rho^A)$, the original defect
profile.  Hence we are guaranteed approximate solutions 
to the field equations near smooth $K^A=0$
(extremal) surfaces.  The argument in this section can be straightforwardly 
extended to gauge fields and so the task of solving the field equations has been
replaced by the task of finding extremal surfaces.

The extrinsic curvature $K^A$ will generically vanish only at $u^A=0$, and be 
non-zero elsewhere in spacetime, and so the static solutions $\bar\Phi$ will 
not be exact. However, we should be able to find approximate solutions
$\Phi=\bar\Phi + \varphi$, with the perturbation $\varphi$ being sourced by the
departures of $K^A$ from zero:
\ben
\varphi(\xi) = \int d^4\xi'\sqrt{-G}\,\De_R(\xi,\xi') K^A(\xi')\pa_A\bar\Phi(u'),
\een
where $\De_R(\xi,\xi')$ is the retarded Green's function for the scalar field fluctuation
operator, given by
\ben
[-^{(d)}\Box +V''(\bar\phi)]\De_R(\xi,\xi') = \de^d(\xi-\xi'),
\een
with $\De_R(\xi,\xi') = 0$ for $\xi^0 < {\xi'}^0$.
If the extrinsic curvature decreases with time, the source for the
perturbation $\varphi$ dies away, and we should not have to worry that our
initial assumption that $|\varphi| \ll |\bar\Phi|$ is rendered invalid.  In
fact, the dynamic scaling hypothesis holds that $K^A \sim \xi^{-1}$, where
$\xi$ is the average curvature radius of the defect network.

There are in fact special cases for which $K^A=0$ everywhere, and exact 
curved defect solutions exist.  These represent travelling waves on walls 
and strings \cite{Vac90}, although they do not obey a superposition principle 
because of the non-linearities in the field theory. 

This brings us close to the controversial subject of radiation from defect 
networks.  We postpone this discussion until Section \ref{s:rad}.

\section{Extremal surfaces}
We saw in the last section that if we could find a suitable surface
of constant $u^A$ (which without loss of generality we can choose to be
$u^A=0$) satisfying $K^A=0$, an approximate solution of the field
equations could be found.
We shall now derive the equations that $u^A$ 
must satisfy in order that $u^A(X)=0$ be an extremal surface.

Differentiating once with respect to the world-volume coordinates
$\sigma^\alpha$, we find
\begin{equation}
\partial_\beta X^\mu \partial_\mu u^A(X) = 0.
\label{eNorm}
\end{equation}
(This equation is of course true independently of the choice of 
the coordinates $\xi$.) 
Using the embedding metric
we can covariantly differentiate (\ref{eNorm}) by acting with 
$(-\gamma)^{-1/2}\partial_\alpha(-\gamma)^{1/2}\gamma^{\alpha\beta}$, 
where $\gamma = \det
\gamma_{\alpha\beta}$, to
obtain
\begin{equation}
^{(p+1)}\Box X^\mu\partial_\mu u^A + \gamma^{\alpha\beta} 
\partial_\alpha X^\mu \partial_\beta X^\nu \partial_\mu\partial_\nu
u^A = 0.
\label{e:ExtEq1}
\end{equation}
The operator 
\ben
^{(p+1)}\Box =
(-\gamma)^{-1/2}\partial_\alpha(-\gamma)^{1/2}\gamma^{\alpha\beta}\pa_\be
\een 
is the covariant d'Alembertian in the surface $u^A=0$.  

The equations of motion are obtained by extremizing the invariant
area of the surface \cite{VilShe94,Hindmarsh:1994re},
\ben
A_{\rm inv}[X] = \int d^{p+1}\si\, \sqrt{-\ga(X)} 
\een
with respect to the enbedding coordinates $X^\mu(\si)$.  The result is
\begin{equation}
^{(p+1)}\Box X^\mu + \Gamma^\mu_{\nu\rho} \gamma^{\alpha\beta} 
\partial_\alpha X^\nu \partial_\beta X^\rho = 0,
\label{e:DEOM}
\end{equation}
where $\Gamma^\mu_{\nu\rho}$ is the affine connection derived from the 
metric $g_{\mu\nu}$. 

The reader will notice the appearance of the tangential 
projector $P_{\parallel}^{\mu\nu}$ in equations (\ref{e:ExtEq1},
\ref{e:DEOM}), which 
we replace by $g^{\mu\nu}-P_{\perp}^{\mu\nu}$
Combining (\ref{e:DEOM}) and (\ref{e:ExtEq1}) to eleminate the d'Alembertian
we find
\begin{equation}
\left[g^{\mu\nu} - h_{AB}\partial^\mu u^A\partial^\nu u^B\right] 
(\partial_\mu\partial_\nu u^C - \Gamma^\rho_{\mu\nu}\partial_\rho 
u^C) = 0.
\label{e:Fund}
\end{equation}
This is the fundamental equation of motion for the fields
$u^A(x)$, which strictly only applies at $u^A=0$.

The equations also follow from a variational procedure. 
Using the fact that $G=\ga/h$, and that $G=g$, one can show that the
invariant area of a $p$-brane can be re-expressed in terms of the $u^A$ 
as
\ben
\A_{\rm inv}[u^A] = \int d^4x \sqrt{-g} \sqrt{h}\de^N(u).
\een
Varying with respect to $u^A$ and dividing by $\sqrt{-\ga}$ gives us 
\ben
\frac{\sqrt{h}}{\sqrt{-g}}\frac{\de \A_{\rm inv}}{\de u^A(x)}
=-\de^N(u)h^{\half}\nabla_\mu
\left(h^{-\half} h_{AB} {\pa^\mu u^B}\right)=0.
\label{e:ExCurDef}
\een
This can be shown to be equivalent to 
\ben
\de^N(u)P_{\parallel}^{\mu\nu}\nabla_\mu \pa_\nu u^A = 0,
\een
and hence Eq.\ (\ref{e:Fund}) at $u^A=0$.
In orthonormal coordinates, for which $h^{AB} = \de_{AB}$, 
Eq.\ (\ref{e:ExCurDef}) becomes
\ben
\de^N(u)\nabla_\mu n^{A\mu} \equiv \de^N(u) K^A= 0,
\een
where $K^A$ is the extrinsic curvature. Thus we can identify ${K_{\mu\nu}^A} = 
P^\si_{\parallel\mu}\nabla_\si\pa_\nu u^A$ as the extrinsic curvature tensor, 
or equivalently the second fundamental tensor (see Table \ref{t:CarterTranslate} and 
\cite{Carter:2000wv}).

The restriction that the equations apply only at $u^A=0$ complicates the finding 
of solutions, 
and we assume that we can extend the equation 
$K^A=0$ to all $u^A$.  It is not 
obvious that non-trivial solutions exist to the extended equations, because such a solution 
would be a 
foliation of space-time in which all leaves have zero extrinsic curvature.  
As mentioned above, some non-trivial solutions are known \cite{Vac90} but there is 
no general existence proof as for the Allen-Cahn equation \cite{MotMeaCur}.
However, 
we could equally well look for solutions to $K^A=f(u^A)$, with $f(u)$ any function which 
vanishes at $u=0$, so there should be a certain amount of freedom.  Furthermore, 
we will be looking only for perturbative solutions to the extended equations.  

\section{Average extremal surfaces}
The equations of motion (\ref{e:Fund}) are not easy to solve, as they are
non-linear.  However, they have the distinct advantages over the 
alternatives.  The equations of motion for the coordinates of the $u^A=0$
surfaces 
(\ref{e:DEOM}) are non-local:  defects
generically self-intersect. This non-locality generally defeats analytic
approaches, and also makes numerical simulations algorithmically 
difficult, as one must devise an 
efficient scheme for searching for self-intersections
\cite{Albrecht:1989mk,StrSimBB,StrSimAS}.  The equations of
motion for the underlying field theory are also non-linear, and in the
gauge of the Abelian Higgs model (and other gauge theories) have a gauge
covariance, which precludes the naive application of techniques like 
large $N$.  Numerical
simulations of field theories are relatively straightforward, but 
require significant amounts memory to allow the scale of the network to
grow much larger than the width of the defect.

Instead of trying to find families of surfaces whose curvature is
exactly zero, we shall find surfaces who curvature is zero {\em on
average}.  The average will be taken with respect to a Gaussian
probability distribution for $u^A$.  We assume that the distribution
function remains Gaussian throughout the evolution, which is similar to
the approximation underlying the large $N$ approximation in scalar field
theory.  Indeed, we should expect there to be a similar large $N$ limit
in this theory.

\subsection{Gaussian averaging}

Our starting point is an ensemble of coordinate functions $u^A(x)$ 
with an assumed Gaussian distribution.  Thus the average value 
of all observables of interest $\Om(u^A,\pa_\mu u^A)$, 
which we take to be
functions of $u^A$ and its derivative  $\pa_\mu u^A$, are evaluated with
probability distribution  \ben
dP[u^A] = \D u \exp \left( -\half \int d^dx d^dy \,u^A(x) C^{-1}_{AB}
(x,y) u^B(y)  \right),
\een
where $C^{AB}(x,y)$ is the 2-point correlation function.

We are often interested in densities, which means that the observable 
$\Om$ is evaluated at a particular point $\tilde x$.  This means we can 
simplify the evaluation of the averages from a functional integral 
to an ordinary one, as we now demonstrate.

First, let us take the Fourier transform of the observable,
\ben
\Om(u^A({\tilde x}),\pa_\mu u^A({\tilde x})) = 
\int \frac{d^N l}{(2\pi)^N}\frac{d^{Nd}k}{(2\pi)^{Nd}} 
\widetilde{\Om}(l,k) e^{i l_A u^A({\tilde x}) + i k^\mu_A \pa_\mu u^A({\tilde x})}.
\een
We now introduce current densities $L_A(x)$ and $K^\mu_A(x)$, 
according to
\ben
L_A(x) = l_A\de^d(x-{\tilde x}), \qquad K^\mu_A(x) = k^\mu_A\de^d(x-{\tilde
x}), 
\label{e:FTFun}
\een
so that the expectation value of $\Om({\tilde x})$ is given by
\ben
\vev{\Om(u^A,\pa_\mu u^A)}
 = 
\int \frac{d^N l}{(2\pi)^N}\frac{d^{Nd}k}{(2\pi)^{Nd}} 
\int dP[u^a] \widetilde{\Om}(l,k) 
e^{ i \int d^dx (L_A(x) - \pa \cdot K_A(x))u^A(x)}.
\een
Performing the integral of the random field $u^A$, we find 
\ben
\vev{\Om(u^A,\pa_\mu u^A)} = 
\int \frac{d^N l}{(2\pi)^N}\frac{d^{Nd}k}{(2\pi)^{Nd}} 
\widetilde{\Om}(l,k) 
e^{ -\half \int_x\int_y \,(L_A - \pa \cdot K_A)
C^{AB}(L_B - \pa \cdot K_B)}.
\een
Substituting 
the form of the functions $L_A$ and $K^\mu_A$ from (\ref{e:FTFun}),
we find
\ben
\vev{\Om} = 
\int \frac{d^N l}{(2\pi)^N}\frac{d^{Nd}k}{(2\pi)^{Nd}} \,
\widetilde{\Om} \,
e^{ -\half l_A C^{AB}(\eta) l_B + l_A[\pa_\mu C^{AB}(\eta)] k^\mu_B -
\half k_A^\mu [\pa_\mu\pa_\nu C^{AB}(\eta)] k^\nu_B},
\een
where
\bea
C^{AB}(\eta) = \lim_{x \to y}C^{AB}(x,y), &\quad&
\pa_\mu C^{AB}(\eta) = \lim_{x \to y}\frac{\pa}{\pa x^\mu} C^{AB}(x,y),
\nonumber\\
\pa_\mu \pa_\nu C^{AB}(\eta) & = & 
   \lim_{x \to y}\frac{\pa}{\pa x^\mu} \frac{\pa}{\pa y^\mu}C^{AB}(x,y),
\eea
and $\eta=\tilde{x}^0$.
We note that we expect correlation functions to be spatially homogeneous 
but to depend on conformal time non-trivially, reflecting the 
symmetries of the background spacetime:  hence the explicit 
conformal time dependence of the two-point correlators evaluated at 
the same two points. 
At this point we recall that the Fourier transform of the observable 
$\Om$ may be written
\ben
\widetilde\Om(l,k) = \int d^Nu d^{Nd}\pi
  \,\Om(u^A,\pi_\mu^A) e^{- il_Au^A - i k^\mu_A \pi_\mu^A}.
\een
We can economise slightly on the length of expressions by introdcuing 
some new notation.  Defining $N(d+1)$-dimensional objects $j$ and 
$f$ by
\ben
j = \{ l_A,k^\mu_A\}, \qquad f = \{ u^A, \pi_\mu^A \},
\een
with a scalar product $(j,f) = l_Au^A + k^\mu_A \pi_\mu^A$, we can 
write
\ben
\vev{\Om(f)} = 
\int d^{N(d+1)}f \,\frac{d^{N(d+1)}j}{(2\pi)^{N(d+1)}} \,
\Om(f) \, e^{-\half (j,\C j) -i (j,f) },
\een
where $\C$ is the covariance matrix
\ben
\C = \left(\begin{array}{cc}
         C^{AB}(\eta) & \pa_\mu C^{AB}(\eta) \\
         \pa_\nu C^{AB}(\eta) & \pa_\mu\pa_\nu C^{AB}(\eta)
        \end{array}\right)
\een
Finally, we may perform the integrations over the Fourier transform 
variables $j$ to obtain
\ben
\vev{\Om(f)} = 
[(2\pi)^{-N(d+1)/2}][\det\C]^{-\half}
\int d^{N(d+1)}f \;
\Om(f) \, e^{-\half (f,\C^{-1} f) }.
\een
Hence the average of the observable can be found with an ordinary integral, 
as claimed.

\subsection{The covariance matrix $\C$}
In our case the covariance matrix $\C$ is restricted by the 
assumed FRW form of the background.  It will be homogeneous 
and isotropic, but not time-independent. We will also assume 
an O($N$) symmetry between the $N$ coordinate functions $u^A(x)$.
Hence, 
the basic two-point correlation function at zero separation may be
written 
\begin{equation}
C^{AB}(\eta) \equiv \langle u^A(x)u^B(x)\rangle = \delta^{AB}
C(\eta),
\end{equation}
We shall also define
a function
$M_{\mu\nu}$ from the two-point correlator of $\partial_\mu u^A$:
\begin{equation}
\pa_\mu\pa_\nu C^{AB}(\eta) \equiv\langle\partial_\mu u^A(x)
\partial_\nu u^B(x)\rangle = \delta^{AB} M_{\mu\nu}(\eta).
\label{e:Mdef}
\end{equation}
The assumed spatial isotropy of the distribution function dictates the
form of $M_{\mu\nu}$:
\begin{equation}
M_{\mu\nu} = \left(\begin{array}{cc}
		T(\eta) & 0 \cr
		0 & \delta_{mn} S(\eta) 
             \end{array}\right).
\end{equation}
With this definition it is not hard to show that $S(\eta) = -C''(\eta)$, where
$C''(\eta) = \lim_{r\to 0}\frac{\pa^2}{\pa r^2}C(\eta,r)$.

Two-point  correlators with odd numbers of derivatives also occur, as the
ensemble is not time translation invariant.  The correlator with
one derivative is
\ben
\pa_\mu C^{AB}(\eta) \equiv \vev{\pa_\mu u^A(x) u^B(x)}
= \half \de^{AB} \de_\mu^0\dot C(\eta),
\een
and with three,
\begin{equation}
\langle \partial_\mu u^A(x)
\partial_\nu\partial_\rho u^B(x)\rangle = 
\gamma_{\mu\nu\rho}(\eta)\delta^{AB}.
\end{equation}
Again, symmetry restricts the form of $\gamma_{\mu\nu\rho}$:
\begin{equation}
\gamma_{000}(\eta) =  \frac{1}{2}\dot T(\eta), \qquad
\gamma_{0mn}(\eta)  =  -\frac{1}{2}\dot S(\eta)\delta_{mn}, \qquad
\gamma_{m0n}(\eta)  =  \frac{1}{2}\dot S(\eta)\delta_{mn}.
\end{equation}
It is interesting to note that $\ga_{\mu\nu\rh} = 
\half(M_{\mu\nu,\rh} + M_{\mu\rh,\nu} - M_{\nu\rh,\mu})$.

Thus the covariance matrix can be written
\ben
\C = \de^{AB} \otimes \left(
       \begin{array}{ccc}
        C & \half \dot C & 0 \\
        \half \dot C & T & 0 \\
        0 & 0 & \de_{mn} S
       \end{array}
      \right)
\een
Its inverse is easily found, and defining the determinant of the 
upper $2\times 2$ block $\De = (TC - \frac{1}{4} \dot C^2)$, we can
write
\ben
\C^{-1} = \de_{AB} \otimes \left(
       \begin{array}{cc}
          \begin{array}{cc}
            T & -\half \dot C \\
            -\half \dot C & C
          \end{array} & 0 \\
        0 & \de_{mn} \frac{\De}{S}
       \end{array}
      \right) \; \frac{1}{\De}
\een
The determinant factor in the probability distribution 
is also straightforward:
\ben
[\det \C]^{-\half} = [S^D \De]^{-N/2}.
\een
Often, we will want to find expectation values which are 
independent of $\pa_0u^A$, mainly because the integrals 
are easier to evaluate.  By integrating over $\pi_0^A$ one 
can easily show that
\bea
\vev{\Om(u^A,\pa_i u^A)} = \hfill &&\nonumber\\ \relax
&\hspace{-3cm}[(2\pi)^{d}S^D C]^{-N/2}
\int d^{N}u d^{ND}\pi\;
\Om(u^A,\pi_i^A) \, e^{-\half u^A\de_{AB}u^B/C-\half\pi_i^A\de_{AB}
\de^{ij}\pi_j^B/S }.&
\eea 
It is very convenient to rescale the integration variables in the probability
distribution, $u^A \to u^A\sqrt{C}$ and $\pi_i^A \to \pi_i^A\sqrt{S}$, in
which case
\bea
\vev{\Om(u^A,\pa_i u^A)} = \hfill &&\nonumber\\ \relax
&\hspace{-3cm}\frac{1}{(2\pi)^{dN/2}}
\int d^{N}u d^{ND}\pi\;
\Om(u^A\sqrt{C},\pi_i^A\sqrt{S})
\, e^{-\half u^A\de_{AB}u^B-\half\pi_i^A\de_{AB}
\de^{ij}\pi_j^B }.&
\label{e:AvgForm}
\eea 

\subsection{Averaging the null extrinsic curvature condition}
The averaging procedure is greatly aided by rewriting the equations of 
motion (\ref{e:Fund}) in the following form:
\begin{equation}
\frac{1}{\sqrt{-g}}
\left[\frac{\partial}{\partial g_{\mu\nu}} \sqrt{-g} \det h \right] 
(\partial_\mu\partial_\nu u^C - \Gamma^\rho_{\mu\nu} \partial_\rho 
u^C) = 0.
\end{equation}
The procedure now is to linearise the equations of motion by 
taking the Gaussian average, and then to find a self-consistent 
solution for the fields $u^A(\bx,\eta)$.  We will require the
following identities, which are proved in Appendix \ref{a:AveExtSurEqn}:
\begin{eqnarray}
\langle \det h\, \partial_\mu\partial_\nu u^C\rangle & = & 
\langle \det h \rangle \partial_\mu\partial_\nu u^C + \frac{2}{N} 
\gamma_{\rho\mu\nu} \left(\frac{\partial}{\partial M_{\rho\sigma}} 
\langle \det h \rangle\right)\partial_\sigma u^C
\label{eExp1}\\
\langle \det h\, \partial_\rho u^C\rangle & = & \langle \det h \rangle
\partial_\rho u^C + \frac{2}{N}\left(\frac{\partial}{\partial 
M_{\eta\sigma}} 
\langle \det h \rangle\right)M_{\eta\rho} \partial_\sigma u^C.
\label{eExp2}
\end{eqnarray}
The expectation value of the determinants in (\ref{eExp1},\ref{eExp2}) can 
be expressed in terms of the two-point correlator $M_{\mu\nu}$ 
(defined in Eq.\ \ref{e:Mdef})
\begin{equation}
\langle \det h \rangle = N! \mathop{\Pi}_{i=1}^N M_{\mu_i\nu_i} 
g^{\rho_i\sigma_i} 
\delta^{\mu_1\ldots\mu_N}_{\rho_1\ldots\rho_N}
\delta^{\nu_1\ldots\nu_N}_{\sigma_1\ldots\sigma_N},
\label{e:ExpDetH}
\end{equation}
where $\delta^{\mu_1\ldots\mu_N}_{\rho_1\ldots\rho_N}$ 
is the identity tensor in the space of rank N 
antisymmetric tensors, defined in Appendix \ref{a:Prelim}.  The right hand 
side of (\ref{e:ExpDetH}) resembles a determinant, and we introduce the
notation $\detbar M$ to refer to it.
We can also define a kind of cofactor for $M_{\mu\nu}$, 
\ben
\overline M^{\mu\nu} = 
N!
\mathop{\Pi}_{i=1}^N 
g^{\rho_i\sigma_i} 
\delta^{\mu\mu_2\ldots\mu_N}_{\rho_1\ldots\rho_N}
\delta^{\nu\nu_2\ldots\nu_N}_{\sigma_1\ldots\sigma_N}
M_{\mu_2\nu_2} \ldots M_{\mu_N\nu_N}/\detbar M.
\een
Putting the pieces together we find that the 
linearised equations for surfaces which
are on average extremal are
\bea
\left(g^{\mu\nu}\right. &-& \left.g^{\mu\rh}g^{\nu\si}
\frac{\pa}{\pa g^{\rh\si}}
\right)
\detbar M \left[
\pa_\mu\pa_\nu u^C\right. + \frac{2}{N} \overline M^{\ka\eta} 
\ga_{\ka\mu\nu} \pa_\ta u^C \nonumber \\
&-& \left.\Ga^\ta_{\mu\nu}\left(\pa_\ta u^C + 
\frac{2}{N} \overline M^{\ka\la}M_{\la\ta}\pa_\la u^C \right)
\right] = 0
\label{e:AvSurf1}
\eea
With the assumed symmetries for the correlation functions, these
equations have the form
\begin{equation}
\ddot u^C + \frac{\mu(\eta)}{\eta} \dot u^C - v^2\nabla^2u^C = 0,
\label{eLEOM}
\end{equation}
where $\mu(\eta)$ and $v$ depend on $T$, $S$, and the background 
cosmology parametrized by $\al$, and must be taken on a case-by-case 
basis for each $N$.

\subsection{Linearised equations for walls and strings}
In our three-dimensional Universe, the cases of most interest are 
$N=1$ (domain walls) and $N=2$ (gauge strings).
$N=3$ corresponds to gauge monopoles, which do not scale 
\cite{KolTur90,VilShe94}. 
For ${N=1}$, $\langle \det h \rangle = 
M_{\mu\nu}g^{\mu\nu}$, while for $N=2$ $\langle \det h \rangle = 
\frac{1}{2} [(M_{\mu\nu}g^{\mu\nu})^2 - M_{\mu\nu}M^{\mu\nu}]$. 
We then find, for FRW backgrounds (see Appendices \ref{a:ueqN1},
\ref{a:ueqN2} )
\begin{equation}
\mu(\eta) = \left\{\begin{array}{ll}
	-2\eta(\dot S /S) + 
	{\alpha(\eta)}\left[ D - 
	3\left({T}/{S}\right)\right] & (N=1),\cr
	-({2}/{(D-1)})\eta({\dot S}/{S}) + 
	{\alpha(\eta)}\left[ (D-1) - 
	4\left({T}/{S}\right)\right] & (N=2),	
	\end{array}\right.
\label{eMuEqn}
\end{equation}
where $\alpha(\eta) = \eta\dot a/a$, and
\begin{equation}
v^2 = \left\{\begin{array}{lr}
	\left[{D-1} - \left({T}/{S}\right) 
	\right]/D & (N=1), \cr
	\left[{D-2} - 2\left({T}/{S}\right) 
	\right]/D & (N=2). 
	\end{array}\right.
\label{eV2Eqn}
\end{equation}
In scaling solutions, we expect $S$ and $T$ to have power law behaviour, and 
so as long as we are not near a transition in the equation of state 
of the Universe (such as that between the radiation- and matter-dominated
eras), $\mu$ and $v^2$ are constant. Thus, imposing the boundary 
condition that $u^C$ be regular as $\eta \to 0$,  
(\ref{eLEOM}) has the simple solution
\begin{equation}
u^C_{\bk}(\eta)  = A^C_{\bk} \left(\frac{\eta}{\eta_{\rm i}}\right)^{(1-
\mu)/2+\nu} \frac{J_\nu(kv\eta)}{(kv\eta)^\nu},
\label{e:LinuEq}
\end{equation}
where $A^C_{\bk} \to 2^\nu \Gamma(\nu+1) u^C_{\bk}(\eta_{\rm i})$ as 
$\bk \to 0$, 
and $(1-\mu)^2/4 = \nu^2$. The form of the initial power spectrum 
is taken to be a power law, with index $q$, and an upper cut-off at $|\bk| =
\La$.

We may now evaluate $T/S$ and $v^2$, and self-consistently 
solve for the undetermined parameter $\mu$. 
It turns out that one must take $\nu = -(1-\mu)/2$ if all the integrals are
required not to diverge as $\La \to \infty$. This also gives regular solutions 
as $\eta \to$, because as it turns out, $\mu > 1$. With this choice, 
$C$ scales as
$\eta^{-(D+q)}$, $S$ and $T$ as 
$\eta^{-(D+q+2)}$.

In the following, we will take the power spectrum to be white noise, $q=0$, as
is consistent with a causal origin for the defects in a phase transition. There
the power spectrum of the scalar field from which the defects are made has a
$q=0$ power spectrum at long wavelengths, and so we should take the fields
$u^A$ to have a similar power spectrum if we want to reproduce the statistics
of the defects from the statistics of the zeros of $u^A$.

Using standard integrals of Bessel functions, 
and defining the parameter $\beta = 2\nu - D - 1 = \mu - D -2$, we find 
(see Appendix \ref{a:Integ}) 
\begin{equation}
\frac{T}{S} =\left\{ \begin{array}{lr} 
 \frac{(D+2)(D-1)}{2(D+2+\beta)}  &(N=1),\cr
 \frac{(D+2)(D-2)}{3(D+2)+2\beta} &(N=2),
 \end{array}\right.
\end{equation}
provided $\beta > 0$, which ensures that the integrals for $S$ and $T$ 
are defined. 
Given the expressions for $T/S$ is easy to show that 
\begin{equation}
v^2 =\left\{ \begin{array}{lr} 
 \frac{(D-1)(D+2+2\be}{2D(D+2+\beta)}  &(N=1),\cr
 \frac{(D-2)(D+2+2\beta)}{D[3(D+2)+2\be]} &(N=2).
 \end{array}\right.
\label{eV2again}
\end{equation}
To find $\beta$, we must solve the equations derived from 
(\ref{eMuEqn}):
\begin{equation}
\be  = \left\{ \begin{array}{lr}
	\alpha\left[D- 3(T/S)\right] + 
	(D+2) & (N=1), \cr
	\alpha \left[D -2 -3(T/S)\right] +(D+2)(3-D)/(D-1) & (N=2), \cr
\end{array}\right.
\label{e:betaeqn}
\end{equation}
which are quickly seen to be quadratic.
One can obtain results in simple closed form in
Minkowski space ($\al=0$) and curvature dominated universes ($\al=\infty$)
which are displayed in Table \ref{t:ExactValuesN1}. For
other backgrounds the
solutions may be written down in closed form, 
but are not
particularly illuminating as they are fairly lengthy expressions.
\begin{table}\begin{center}
\begin{tabular}{|c|c|ccc|}
\hline
N & $\alpha$ & $\beta$ & $(T/S)$ & $v^2$ \\
\hline
1 &  0 & 
$(D+2)$ & 
        $\frac{D-1}{4}$  &  
                $ \frac{3}{4}\frac{D-1}{4}$  \\[1pt]
 & $\infty$ & 
$ \frac{(D+2)(D-3)}{2D}$ & 
        $ \frac{D}{3}$ &
                $ \frac{2D-3}{3D}$ \\[1pt]
\hline
2 &  0 & 
$ \frac{(D+2)(3-D)}{D-1}$ & 
        $ \frac{(D-2)(D-1)}{D+3}$ &  
                $ \frac{(D-2)(D+5)}{D(D+3)}$ \\[1pt]
 &  $\infty$ & 
        0 & $\frac{D-2}{3}$   & 
                $ \frac{D-2}{3D}$   \\[1pt]
\hline
\end{tabular}
\caption{
Values for parameters $\beta$, $v^2$ and $T/S$ of the self-consistent
solution to the linearised equations of motion (\ref{eLEOM})
for the $N$ fields $u^A$ for $N=1$ 
(domain walls). In the special cases of  Minkowksi space $(\alpha=0)$, and 
curvature-dominated FRW cosmologies ($\alpha=\infty$), exact values can be 
found for all $D$.
\label{t:ExactValuesN1}}
\end{center}\end{table}

Instead, numerical values of $\be$, $T/S$ and $v^2$ for particular
cases of interest are given: radiation-dominated 
$(\alpha =1)$ and
matter-dominated $(\alpha=2)$ 2 and 3-dimensional universes 
(Tables
\ref{tValuesN1},
\ref{tValuesN2}).

Note that for strings ($N=2$) in 3 dimensions
in Minkowski space ($\al=0$), for which $\mu = D+2$, and hence $\beta=0$,
which does not satisfy the requirement $\beta>0$ for the 
integrals defining $S(\eta)$ and $T(\eta)$ to be convergent. One finds that 
a logarithmic scaling violation appears, and 
$S,T \propto \log(\La\eta)\eta^{-(D+2+q)}$. 
We also have a solution with $\be=0$, and therefore logarithmically
divergent $S$ and $T$, for walls in 3-dimensional curvature-dominated
universes.

\begin{table}\begin{center}
\begin{tabular}{|c|ccc|ccc|}
\hline
 & & $D=3$ & & & $D=2$ & \\
\hline
$\alpha$ & $\beta$ & $(T/S)$ & $v^2$ & $\beta$ & $(T/S)$ & $v^2$ \\
\hline
0        & 5    & 1/2  & 1/2  &  4    & 1/4   & 3/8   \\
1        & 6.72 & 0.43 & 0.52 &  5.36 & 0.21  & 0.39  \\
2        & 8.83 & 0.36 & 0.55 &  6.90 & 0.18  & 0.41  \\
$\infty$ & 0    & 1    & 1/3  &  -1   & 2/3   & 1/6   \\
\hline
\end{tabular}
\caption{
Values for parameters $\beta$, $v^2$ and $T/S$ of the self-consistent
solution to the linearised equations of motion (\ref{eLEOM})
for the $N$ fields $u^A$ for $N=1$ 
(domain walls) in $D=2,3$. Values listed 
are for Minkowksi space $(\alpha=0)$, 
radiation-dominated $(\alpha=1)$,
matter-dominated $(\alpha=2)$,and 
curvature-dominated FRW cosmologies ($\alpha=\infty$).
\label{tValuesN1}}
\end{center}\end{table}

\begin{table}\centering
\begin{tabular}{|c|ccc|}
\hline
 & & $D=3$ & \\
\hline
$\alpha$ & $(T/S)$ & $v^2$ & $\be$ \\
\hline
0        & 1/3  & 1/9  &  0    \\
1        & 0.22 & 0.14 &  3.65 \\
2        & 0.20 & 0.20 &  4.75 \\
$\infty$ & 1/3  & 1/9  &  0    \\
\hline
\end{tabular}
\caption{
Values for parameters $\beta$, $v^2$ and $T/S$ of the self-consistent
solution to the linearised equations of motion (\ref{eLEOM})
for the $N$ fields $u^A$ for $N=2$ in $D=3$ 
(strings). Values listed 
are for Minkowksi space $(\alpha=0)$, radiation-dominated $(\alpha=1)$,
matter-dominated $(\alpha=2)$,
and curvature-dominated FRW cosmologies ($\alpha=\infty$).
\label{tValuesN2}}
\end{table}

\section{Area densities for walls and strings}
Armed with the mean-field solution for $u^A(x)$ we can 
now
calculate anything that can be expressed in terms of local functions of 
the field and its derivatives, provided of course that we are able to
perform the Gaussian integrals involved.  Here we derive formulae for
the area densities of defects, where by ``area'' we mean the world-volume of
the $(p+1)$-dimensional hypersurface $u^A=0$, which has dimensions of $({\rm
Length})^{-N}$.  We must be careful to
distinguish between various kinds of area: there is invariant or proper area
which is a coordinate-independent quantity, and there is also the
projected
$p$-dimensional area. The latter quantity is what one would
obtain by simply measuring the $p$-dimensional area of the defects at a
particular time.  This quantity is the most convenient to calculate for
comparison with numerical simulations, which is a good thing as the proper area
density is far harder to calculate.  One must also bear in mind that area
{\em densities} are coordinate-dependent quantities: in the cosmological
setting we will need
to convert between comoving area density and physical area density by
multiplying by the appropriate power of the scale factor $a$, which is 
$a^{-N}$.

Here we give figures for the projected area 
 densities of walls and strings in $D=3$. They can be compared with results
from numerical simulations of the field theories 
and give surprisingly good agreement given the uncontrolled nature of the
approximations made.
 
\subsection{Proper area density}

The proper area density $\A$ of a $p$-dimensional defect in $D$ space 
dimensions is 
\begin{equation}
\A_D^p(x) = \int d^{p+1}\sigma'\,\sqrt{-\gamma} \delta^{d} 
(x-X(\sigma'))/\sqrt{-g}.
\end{equation}
Making the coordinate transformation from $x^\mu$ to $\xi^\mu =
\{\sigma^\alpha, u^A\}$ near the world volume of the defect, we have 
\begin{equation}
\A_D^p(\xi) =  \int d^{p+1}\sigma'\,\sqrt{-\gamma} 
      \de^{p+1}(\si-\si')\delta^N(u^A)/\sqrt{-G},
\end{equation}
Recalling the results of Section (\ref{s:CurvCoord}), we can perform the
integration over $\si'$, to obtain
\begin{equation}
\A_D^p =  \delta^N(u^A)|\det h^{AB}|^{1/2},
\label{e:WVDen}
\end{equation}
where the reader is reminded that
\ben
h^{AB} = \pa_\mu u^A \pa_\nu u^B g^{\mu\nu}
\een
Thus the problem of calculating the proper area density is
reduced to finding the Gaussian average 
of $\A_D^p$ in (\ref{e:WVDen}). The conversion factor from comoving to physical
area is given as
\ben
\A_{D,{\rm phys}}^p = a^{-N}\A_D^p,
\een
with $N=D-p$.

\subsection{Projected area density}

Easier to measure and to calculate is the projected area density,  which is
defined as
\ben
A_D^p = \int d^{p}\sigma'\,\sqrt{\gamma_D} \delta^{D} 
(\bx-\bX(\sigma'))/\sqrt{g_D},
\een
where $g_{Dij}$ is the spatial part of the metric.  The induced $D$-dimensional
metric on the $p$-dimensional surface $u^A=0$ is  
\ben
\ga_{Dab} = \pa_a X^i \pa_b X^j g_{Dij},
\een
where $a,b=1,\ldots,p$.  As for the proper area density, one can show that
\ben
A_D^p =  \delta^N(u^A)|\det h_D^{AB}|^{1/2},
\label{e:ProjDen}
\een
where
\ben
h^{AB} = \pa_i u^A \pa_j u^B g_D^{ij}.
\een
Note that $g_D^{ij}$ is defined as the matrix inverse of $g_{Dij}$, 
and is not the spatial part of $g^{\mu\nu}$.
The conversion between physical and comoving area is again
\ben
A_{D,{\rm phys}}^p = a^{-N}A_D^p.
\een

\subsection{Average projected area density: walls}
\label{s:AvPAD1}
We can now use the averaging formula (\ref{e:AvgForm}) to find the mean value
of the operator $A$, which when we specialise to domain walls ($N=1$) gives
\ben
\vev{A_D^{D-1}} = \frac{1}{(2\pi)^{d/2}}\sqrt{\frac{S}{C}}
\int du d^{D}\pi\;
\de(u)|\pi_i|
\, e^{-\half u^2-\half\pi_i
\de^{ij}\pi_j }.
\een
The integrals are easily performed to give
\ben
\vev{A_D^{D-1}} = \sqrt{\frac{S}{\pi C}}\frac{\Ga[(D+1)/2]}{\Ga(D/2)},
\een
a well-known result originally derived by Ohta, Jasnow and Kawasaki
\cite{OhtJasKaw82}.  This is the {\em comoving} projected area
density: to obtain the
physical projected area density, one multiplies by $a^{-1}$:

\subsection{Average projected area density: strings}
\label{s:AvPAD2}
For strings ($N=2$), the average we need to calculate is
\ben
\vev{A_D^{D-2}} = \frac{1}{(2\pi)^{d}}{\frac{S}{C}}
\int d^2u d^{2D}\pi\;
\de^2(u^A)|h^{AB}|^{\half}
\, e^{-\half u^A\de_{AB}u^B-\half\pi_i^A
\de^{ij}\de_{AB}\pi_j^B },
\een
where the rescaled quantity $h^{AB}$ is given by
\ben
h^{AB} = \pi_i^A \pi_i^B.
\een
Now,
\bea
\det h^{AB} &=& \half \ep_{AC}\ep_{BD} h^{AB} h^{CD}, \\
 &=& \half \ep_{AC}\ep_{BD}\pi_i^A \pi_j^C \pi_i^B \pi_j^D,
\eea
which suggests that we construct the following antisymmetric
matrix:
\ben
f_{ij} = \pi_i^A \pi_j^B \ep_{AB},
\een
such that
\ben
\det h^{AB} = \half f_{ij}f_{ij}.
\een
Thus in order to calculate the average area, we need the probability
distribution for $f_{ij}$.  At this point we specialise to $D=3$, as the
calculations are considerably simplified by introducing the vector
\ben
\phi_k = \half \ep_{ijk}f_{ij},
\een
whereupon
\ben
\det h^{AB} = |\phi_k\phi_k|^{\half}.
\een
The probability distribution for $\phi=|\phi_k\phi_k|^{\half}$
is derived in Appendix \ref{a:ProDisF},
and turns out to be remarkably simple, giving
\ben
\vev{A_3^1} = \frac{1}{2\pi}{\frac{S}{C}}
\int d^2u^A d^3\phi\;
\de^2(u^A)\phi
\, e^{-\half u^Au^B\de_{AB}}\frac{1}{4\pi\phi}e^{-\phi}.
\een
A simple calculation now shows that the comoving projected length density for
strings in $D=3$ is 
\ben
\vev{A_3^1} = \frac{S}{\pi C}.
\een
Note that this disagrees with the formula derived by Toyoki and Honda
\cite{ToyHon87}, but agrees with Scherrer and Vilenkin \cite{Scherrer:1997sq}. 
Toyoki and Honda write the 3D string length density as
\ben
A_3^1 = \de(u^1)\de(u^2)|\nabla u^1 \times \nabla u^2| = 
\de(u^1)\de(u^2)|\nabla u^1||\nabla u^2|\cos\theta_{12}
\een
where $\theta_{12}$ is the angle between the vectors $\nabla u^1$ and 
$\nabla u^2$.  They then average $\theta_{12}$ over a uniform distribution,
separately from $u^1$ and $u^2$, which is incorrect.

\subsection{Projected area density: higher $N$}
\label{s:ProAreDenHighN}
Scherrer and Vilenkin \cite{Scherrer:1997sq} used an elegant argument to derive their
value for the projected area densities of walls, strings and monopoles in
$D=3$, which can be generalised to any $N$ and $D$.  They noted that a
string was located at the intersection of two surfaces $u^1=0$ and $u^2=0$, and
therefore the length
density string could found by computing the length per unit area of the lines
of $u^2=0$ in the surface $u^1=0$, and then multiplying by the area per unit
volume the surface $u^1=0$.  That is,
\ben
A_3^1 = A_{2}^1A_{3}^2,
\een
which clearly has the correct dimensions.  One can easily check that this gives
the correct result $A_3^1= (S/\pi C)$.  It is immediately obvious how to
generalise the formula to any $D$ and $N$:
\ben
A_D^p = \prod_{n=p}^{D-1} A_{n+1}^n.
\een
Thus
\ben
A_D^p = \left(\frac{S}{\pi C}\right)^{N/2} \frac{\Ga[(D+1)/2]}{\Ga[(D-N+1)/2]}.
\een
where $N=D-p$,


\subsection{Quantitative results}

It is shown in Appendix (\ref{a:Integ}) that
\begin{equation}
\frac{S}{C} = \frac{1}{\eta^2}\frac{D+2+\be}{4 v^2}\frac{\beta+1}{\beta},
\label{eSoC}
\end{equation}
In the special cases of $N=1,2$ one can substitute for $v^2$ from Eq.\
(\ref{eV2again}) to obtain
\begin{equation}
\frac{S}{C} =\left\{ \begin{array}{lr} 
 \frac{1}{\eta^2}\frac{D(D+2+\be)(1+\be)}{2(D-1)\beta}  &(N=1),\cr
 \frac{D[3(D+2)+2\be](1+\be)}{4(D-2)\be} &(N=2),
 \end{array}\right.
\end{equation}
It was shown in Section \ref{s:ProAreDenHighN} that the projected 
area density is proportional to $(S/C)^{N/2}$, and therefore 
classical scaling behaviour for all defects is predicted, unless $\be = 0$.  By
classical scaling, we mean that the area density goes in
proportion to conformal time as naive dimensional analysis would predict: a
$p$-dimensional area density in $D$ dimensions should be proportional to
$\eta^{-N}$, as indeed it is in this theory.  When $\be=0$, as is the case 
for $(D-2)$-branes in $D=3$ (strings) in Minkowski space, and for 
$(D-1)$-branes in curvature-dominated FRW backgrounds, logarithmic 
violations to naive scaling appear.

\begin{table}
\begin{center}
\begin{tabular}{|c|c|c||c|c|}
\hline
\multicolumn{3}{|c||}{{D=3}
\( A_3^2 = \frac{2}{\pi}\sqrt{\frac{S}{C}}\)} &
\multicolumn{2}{c|}{{D=2}  \( A_2^1 = \frac{1}{2}\sqrt{\frac{S}{C}}\)} \\
\hline
$\alpha$ & \textit{Theory}
& \textit{Simulation} & \textit{Theory} & \textit{Simulation}\\
\hline
0 & \(1.91 \eta^{-1}\)  & {$0.88(0.14). \eta^{-1.00(0.03)}$} 
  &  {$1.11 \eta^{-1}$}  & {$0.77(0.23).\eta^{-0.99(0.03)}$} \\
1 & \(2.02 \eta^{-1}\)  & {$0.93(0.13).\eta^{-0.99(0.01)}$} 
  & {$1.18 \eta^{-1}$}  & {$0.93(0.17).\eta^{-1.00(0.02)}$} \\  
2 & \(2.16 \eta^{-1}\)  & {$0.96(0.12).\eta^{-1.00(0.01)}$} 
  & {$1.24 \eta^{-1}$}  & {$1.15(0.23).\eta^{-0.99(0.01)}$} \\
\hline
\end{tabular}
\caption{
Comparison between theoretical and numerical simulation 
values of the domain wall defect scaling density in Minkowski 
space, FRW radiation and 
FRW matter
dominated universes ($\alpha=0,1,2$  respectively) in 2 and 3 dimensions.
The numerical values are taken from \cite{GarHin02}.
\label{tDenN1}}
\end{center}
\end{table}

We are also able to compute the scaling \textit{amplitudes}, the 
coefficients of the relations between the area density and the appropriate
power of time.  These can then be compared with 
numerical simulations.  
The scaling projected comoving area densities
for walls and strings, in 
the radiation and matter eras are displayed in Tables \ref{tDenN1}, 
\ref{tDenN2}.  Note that in Table \ref{tDenN2}, the results for strings in 
matter and radiation-dominated universes have ben taken from 
\cite{Moore:2001px}, who give {\em proper} area densities.  These 
have been converted to projected area densities by dividing by 
$\vev{(1-v^2)^{-1/2}}$, where $v$ is the average speed of the string.  
While not strictly the correct procedure, it gives a good enough answer 
given the uncertainty.

\begin{table}

\begin{center}
\begin{tabular}{|c|c|c|}
\hline
\multicolumn{3}{|c|}{{D=3}
\( A_3^1 = \frac{S}{\pi C}\)} \\
\hline
$\alpha$ & \textit{Theory} & \textit{Simulation}
\\
\hline
0 & \(3.6 \eta^{-2}\log(\eta\La)\)  & {$(11\pm 1) \eta^{-2}$} \\
1 & \(6.8 \eta^{-2}\)               & {$(18\pm 6) \eta^{-2}$} \\
2 & \(7.1 \eta^{-2}\)               & {$(14\pm 4) \eta^{-2}$} \\
\hline
\end{tabular}
\end{center}
\caption{
Comparison between theoretical and numerical simulation 
values of the string scaling density in Minkowski 
space, FRW radiation and FRW matter
dominated universes ($\alpha=0,1,2$  respectively) in 3 dimensions.
The numerical values are taken from \cite{Vincent:1998cx} and 
\cite{Moore:2001px}, with the latter converted from proper to 
projected area densities.  The numerical fits in Minkowski space did not 
look for a logarithmic scaling violation.
\label{tDenN2}}
\end{table}

To convert between comoving and physical areas, one uses the formula
$A_{\rm phys}(t) = a^{-N}A(\eta)$, and the fact that $a(\eta)\eta = (1+\al)t$.

The scaling amplitudes differ from those obtained in numerical simulations 
of $\phi^4$ theory \cite{GarHin02} and of the Abelian Higgs model 
\cite{Vincent:1998cx,Moore:2001px}, by a factor of about $2$.  However, it
should be noted that there are large errors on the central value.  The authors
of Refs.\ \cite{Vincent:1998cx,Moore:2001px} did not look for logarithmic
scaling violations in the area density for strings in Minkowski space,
choosing instead to fit to a simple 
power law.  Finding such a violation is numerically very demanding, as a large
dynamic range is required.

\subsection{Biased initial conditions}
One may also ask how the network behaves when a small 
bias is introduced into the initial conditions, that is, if 
$\langle u^A(x_i) \rangle = U^A$.  In numerical 
experiments simulating biased initial conditions for strings 
\cite{Perc} it is found that as the bias is increased 
the string passes through a transition from a phase with  
a finite fraction of percolating ``infinite'' string and with 
a power-law size distribution of loops, 
to one without infinite string, and with an exponential size 
distribution for the loops.  In numerical simulations of domain 
walls \cite{Coulson:1995nv,Larsson:1996sp}, 
it is found that even for very small initial 
biases, for which the walls percolate, the system still evolves 
away from the percolating state and eventually the large walls 
break up and disappear. 
Similar behaviour is well-known in 
in the study of quenches of condensed matter systems with a
non-conserved order parameter 
\cite{OhtJasKaw82,ToyHon86,MonGol92,FilBraPur95}. 

The theoretical description of this behaviour is fairly straightforward.
Introducing a bias into the initial conditions  for walls alters the Gaussian
average of Section \ref{s:AvPAD1} to
\ben
\vev{A_D^1} = \frac{1}{(2\pi)^{d/2}}\sqrt{\frac{S}{C}}
\int du d^{D}\pi\;
\de(u)|\pi_i|
\, e^{-\half (u-U/\sqrt{C})^2-\half\pi_i
\de^{ij}\pi_j },
\een
and hence 
\ben
\vev{A_D^1} = \sqrt{\frac{S}{\pi C}}\frac{\Ga[(D+1)/2]}{\Ga(D/2)}
e^{-\half U^2/C}.
\label{e:BiasDen}
\een
It is clear that this form is common to all defects in all dimensions: if
$\vev{A_D^1}_0$
is the unbiased average area density, then the result of including
a bias is
\ben
\vev{A_D^1} = \vev{A_D^1}_0e^{-\half U^2/C},
\een
with an obvious generalisation to $N>1$.
If the system is close to being self-similar at some initial time
$\eta_{\rm i}$ when the magnitude of the bias is $U$ and the fluctuation
around that value $C(\eta_{\rm i})$, then one can predict 
that the area density goes as
\ben
\A \sim \eta^{-N/2} \exp(-cU^2\eta^D),
\een
where $c$ is a constant.
One can also show that the time 
$\eta_{\rm c}$ at which the defect density falls to a fraction $e^{-1}$
of its scaling value as 
\begin{equation}
\eta_{\rm c} = \eta_{\rm i}\left(U^2/2C(\eta_{\rm i})\right)^{-1/D}.
\label{e:DisTime}
\end{equation}
The simulations by Larsson and White are consistent with 
(\ref{e:BiasDen}) and (\ref{e:DisTime}) in $D=2$, but do not have 
sufficiently good statistics in $D=3$ \cite{Larsson:1996sp}.  Coulson 
et al.~\cite{Coulson:1995nv} did not attempt a fit of the form 
(\ref{e:BiasDen}) to their simulations.

\section{Scaling and energy loss}
\label{s:rad}
There is an apparent inconsistency in our conclusions 
for topological defect networks.
We started by establishing that one could find approximate solutions to the 
field equations by finding extremal surfaces in spacetime, and then 
constructing static solutions in coordinates which moved with the surface. 
We then showed that one could construct random surfaces in 
FRW spacetimes which are on average extremal, whose average area density 
obeyed a classical scaling law with conformal time $\eta$.  The assumption is 
that there are defect-like solutions which are somehow close to static 
solutions centred on these random surfaces.

There is a problem with this picture: the defect area density decreases
with time and therefore the energy in the form of defects also decreases.
This energy must go somewhere, and an obvious channel is into propagating
modes of the fields, or radiation. However, it is difficult to reconcile
the idea that the network energy is lost into radiation with the
perturbative approach to finding curved defect solutions, which assumes
that the deviation from the comoving static solution decreases with the
curvature of the defect.

Indeed, there is good numerical evidence that the perturbative approach
works in certain cases \cite{Moore:2001px,Olum:1999sg}. The configurations where it
has been tested are colliding travelling waves, either sinusoidal
\cite{Moore:2001px} or more complex \cite{Olum:1999sg}. When travelling waves are
correctly prepared to the recipe laid down by Vachaspati \cite{Vac90}, the
collision does produce perturbations in the form of radiation, which is
however exponentially suppressed with decreasing curvature.

It should be noted however, that pure travelling waves are obtained from
very special initial conditions. A random defect network is not prepared so
carefully and it appears that it does radiate by an as yet poorly
understood mechanism \cite{Vincent:1998cx,Moore:2001px}. The radiation shows
no sign of being exponentially suppressed with increasing curvature. What
is clear is that one or more of the assumptions implicit in the
perturbative approach to finding curved defect solutions must be violated.
Two possibilities are that the extrinsic curvature is much larger than
$\xi^{-1}$, maybe due to kinks, or there are non-linear radiative
processes, perhaps involving the breather modes \cite{Arodz:1995pt}.

\section{Summary and conclusions}
To summarise, this paper describes a new analytic technique for describing the 
dynamics of a random network of branes or topological
defects, applicable to the brane gas universe or a cosmological phase transition.
It is a relativistic version of a well-known approach in condensed
matter physics, due to Ohta, Jasnow and Kawasaki \cite{OhtJasKaw82}, 
which uses a mean field approach to find approximate solutions to the Allen-Cahn
equation for the motion of a surface representing a phase boundary.  
In the relativistic version, the surfaces are branes or defects obeying the Nambu-Goto
equation (i.e.\ they have zero extrinsic curvature), but the condensed matter
analogues can be obtained as a certain limit (see Appendices \ref{a:ueqN1} and
\ref{a:ueqN2}), which acts as a check. 
In rederiving these condensed matter results an expression for the length density of 
strings due to Toyoki and Honda \cite{ToyHon87} has been corrected (see Section \ref{s:AvPAD2}).

In most cases the
prediction is that the (generalised)
area density of a $p$-dimensional defect in $D$ dimensions should scale with
conformal time as $\eta^{-(D-p)}$, with a scaling amplitude of $O(1)$. This
appears to agree quantitatively with numerical simulations of domain walls 
\cite{Larsson:1996sp,GarHin02}.
In certain cases, such as strings in $D=3$, there is a
prediction of a logarithmic violation of the naive scaling law.
There are further predictions for defects with biased initial conditions, 
for strings in 3D, and for $(D-1)$- and $(D-2)$-branes which would be
interesting to test.  

From the point of view of the brane gas Universe, it
would be interesting to look at 1-, 2- and 5- branes in higher dimensions.  One
of the most interesting features of the brane gas scenario is that it offers
and explanation of why the Universe has three large dimensions: strings do not
generically interact with each other in more than three dimensions, and so
winding modes can never decay.  It is only a 3-dimensional subspace, where
the winding modes can interact with each other and annihilate, which can expand
and becaome large.  It follows from this idea that strings cannot scale in more
than 3 space dimensions, as there is no opportunity for the initial winding
modes to break up into closed loops in the conventional picture of energy loss
by a string network.  It is therefore important to see whether the
theoretical techniques presented in this paper predict scaling for strings in
higher dimensions.
\begin{acknowledgments}
I am extremely grateful to Alan Bray,  
for inspiration and for collaboration on the work of Section 
\ref{s:AvPAD2}, and to Ed Copeland and Ian Moss for useful discussions.
I am also pleased to acknowledge support from the CERN Theory Division where
some of this work was done.
\end{acknowledgments}

\appendix

\section{Preliminaries}
\label{a:Prelim}
Define a projector onto the rank $N$ antisymmetric tensors 
(which is also an identity operator for those tensors).
\ben
\de_{\mu_1\ldots\mu_N}^{\nu_1\ldots\nu_N}
= \frac{1}{N!}\left( \de_{\mu_1}^{\nu_1}\ldots\de_{\mu_N}^{\nu_N} 
+ \textrm{signed perms on } \nu_i \right)
\een
This projector has the properties
\bea
\de_{\mu_1\ldots\mu_N}^{\nu_1\ldots\nu_N} &= &
\frac{1}{N!(d-N)!} \ep_{\mu_1\ldots\mu_N\mu_{N+1}\ldots\mu_d}
\ep^{\nu_1\ldots\nu_N\mu_{N+1}\ldots\mu_d},\\
\de_{\mu_1\ldots\mu_N}^{\nu_1\ldots\nu_N}
\de_{\nu_1\ldots\nu_N}^{\rho_1\ldots\rho_N} &= &
\de_{\mu_1\ldots\mu_N}^{\rho_1\ldots\rho_N},\\
\de_{\mu_1\ldots\mu_N}^{\mu_1\ldots\mu_N} &=& \frac{d!}{N!(d-N)!}
\eea
Define the matrix $h^{AB} = \pa_\mu u^A\pa_\nu u^Bg^{\mu\nu}$. Then 
\ben
\det h = \frac{1}{N!}
g^{\mu_1\nu_1}\ldots g^{\mu_N\nu_N}
\pa_{\mu_1}u^{A_1}\pa_{\nu_1}u^{B_1}\ldots 
\pa_{\mu_N}u^{A_N}\pa_{\nu_N}u^{B_N}
\ep_{A_1\ldots A_N}\ep_{B_1\ldots B_N}
\een
Define the antisymmetric rank $N$ tensor
\ben
F_{\mu_1\ldots\mu_N} = \pa_{\mu_1}u^{A_1}\ldots\pa_{\mu_N}u^{A_N}
\ep_{A_1\ldots A_N}.
\een
Then we may write
\ben
\det h = \frac{1}{N!} F_{\mu_1\ldots\nu_N}F^{\nu_1\ldots\nu_N}
\een
Note that
\bea
\frac{\pa}{\pa g^{\mu\nu}} \det h &=&
\frac{\pa_\mu u^{A_1} \pa_\nu u^{B_1}}{(N-1)!}
\left(\pa u^{A_2}\cdot\pa u^{B_2}\ldots
\pa u^{A_N}\cdot\pa u^{B_N}
\ep_{A_1\ldots A_N}\ep_{B_1\ldots B_N} \right)\nonumber\\
 &=&\pa_\mu u^{A_1} \pa_\nu u^{B_1} \cdot h_{A_1B_1} \det h.
\label{e:detID1}
\eea

\section{Averaging the extremal surface equation}
\label{a:AveExtSurEqn}
In a general spacetime, the equation for a $D$-dimensional 
surface with zero extrinsic curvature is 
\ben
\left(g^{\mu\nu} - h_{AB} \pa^\mu u^A \pa^\nu u^B \right)
(\pa_\mu\pa_\nu u^C - \Ga_{\mu\nu}^\ta\pa_\ta u^C)=0,
\een
where 
\bea
h_{AB} &=& (h^{AB})^{-1} \nonumber\\
&=& \frac{1}{(N-1)!}\frac{1}{\det h}
\left(
\pa u^{A_2}\cdot\pa u^{B_2}\ldots
\pa u^{A_N}\cdot\pa u^{B_N}
\ep_{A_1\ldots A_N}\ep_{B_1\ldots B_N}\right)\nonumber
\eea
The surfaces of constant $u^C$ satisfying this equation have 
$K^C=0$. Note that the following is the projector onto the 
tangent space of the surface of constant $u^C$:
\ben
P_\parallel^{\mu\nu} = g^{\mu\nu} - h_{AB} \pa^\mu u^A \pa^\nu u^B 
\een
Thus, if we write $v^C = \pa u^C$ as the coordinate vectors 
normal to the surfaces of constant $u^C$, we can express 
the equation as
\ben
P_{\parallel}^{\mu\nu}\nabla_\mu v_\nu^C = 0.
\een

Recalling the identity \ref{e:detID1} we see that  
the following equation holds:
\bea
\left( g^{\mu\nu} - 
 \frac{\pa}{\pa g_{\mu\nu}}
 \right)
\left[\det h \left( \pa_\mu\pa_\nu u^C - \Ga_{\mu\nu}^\ta\pa_\ta u^C\right)
\right] = 0
\eea
Hence, in order to obtain the equations for surfaces whose 
{\it average} extrinsic curvature is zero, we need to average 
the quantities
$\det h \pa_\mu\pa_\nu u^C$ and $\det h \pa_\ta u^C.$

\subsection{The Gaussian average $\langle\det h\, \pa_\mu\pa_\nu u^C \rangle$}

Exploiting its antisymmetry, 
we may rewrite the tensor $F_{\mu_1\ldots\mu_N}$ as
\ben
F_{\mu_1\ldots\mu_N} = N!\de_{\mu_1\ldots\mu_N}^{\nu_1\ldots\nu_N} 
\pa_{\nu_1}u^1\ldots\pa_{\nu_N}u^N.
\een
Hence the determinant becomes
\ben
\det h = N! \de_{\mu_1\ldots\mu_N}^{\nu_1\ldots\nu_N}
\pa^{\mu_1}u^1\pa_{\nu_1}u^1\ldots\pa^{\mu_N}u^N\pa_{\nu_N}u^N.
\een
We introduce $m^A_{\mu\nu} = \pa_\mu u^A \pa_\nu u^A$ (with no implied 
summation), which is an unnormalised projector orthogonal to the surfaces 
of constant $u^A$.  Then
\ben
\det h = N!\de^{\mu_1\ldots\mu_N\nu_1\ldots\nu_N} 
m^{1}_{\mu_1\nu_1} \ldots m^{N}_{\mu_N\nu_N}.
\een
Hence
\bea
\langle \det h \pa_\mu\pa_\nu u^C \rangle &=& 
\langle \det h \rangle \pa_\mu\pa_\nu u^C\nonumber\\ 
&\hspace{-40mm}+& 
\hspace{-20mm}N! \de^{\mu_1\ldots\mu_N\nu_1\ldots\nu_N} \langle
m^{1}_{\mu_1\mu_2}\ldots
\widehat{m^{C}_{\mu_C\nu_C}} 
\ldots m^{N}_{\mu_N\nu_N}\rangle
 \langle m^{C}_{\mu_C\nu_C}  \pa_\mu \pa_\nu u^C \rangle,\nonumber
\eea
with no implied summation on the index $C$, and the wide hat symbol is used to 
denote a term removed from the product inside the angle brackets.
We now use the relations
\bea
\vev{\pa_{\mu}u^A\pa_{\nu}u^B} &=& M_{\mu\nu}\de^{AB},\\
\vev{\pa_{\rh}u^A\pa_{\mu}\pa_{\nu}u^B} &=& 
\ga_{\rh\mu\nu}\de^{AB},\\
\langle m^{C}_{\mu_1\nu_1}  \pa_\mu \pa_\nu u^C \rangle 
&=& \ga_{\mu_1\mu\nu}\pa_{\nu_1}u^{C}
+ \ga_{\nu_1\mu\nu}\pa_{\mu_1}u^{C}
\eea
from which we can immediately derive
\ben
\vev{\det h} = N! \de^{\mu_1\ldots\mu_N\nu_1\ldots\nu_N} 
M_{\mu_1\nu_1} \ldots M_{\mu_N\nu_N}
\een
and
\bea
N! \de^{\mu_1\ldots\mu_N\nu_1\ldots\nu_N}
\langle m^{1}_{\mu_1\nu_1} \ldots 
\widehat{m^{C}_{\mu_C\nu_C}} \ldots m^{N}_{\mu_N\nu_N}\rangle
\langle m^{C}_{\mu_C\nu_C}  \pa_\mu \pa_\nu u^C \rangle
&=&\nonumber\\ 
\frac{2}{N} \frac{\pa}{\pa M_{\rh\si}} \vev{\det h} 
\ga_{\rh\mu\nu}\pa_{\si}u^C & & 
\eea

\subsection{The Gaussian average $\langle\det h \pa_\ta u^C \rangle$}

It follows from the previous section that 
\bea
&&\hspace{-5mm}\langle\det h \pa_\ta u^C \rangle = 
\vev{\det h} \pa_\ta u^C \nonumber\\ 
&&+
2N! \de^{\mu_1\ldots\mu_N\nu_1\ldots\nu_N}
M_{\mu_1\nu_1}\ldots\widehat{M_{\mu_C\nu_C}}\ldots
M_{\mu_N\nu_N} M_{\mu_C\ta} \pa_{\nu_C}u^C.
\eea

\subsection{More definitions}
We define a kind of determinant $\detbar$ through the 
relation
\ben
\detbar M = N! \de^{\mu_1\ldots\mu_N\nu_1\ldots\nu_N} 
M_{\mu_1\nu_1} \ldots M_{\mu_N\nu_N} = \vev{\det h}.
\een
We can therefore define a cofactor for $M_{\mu\nu}$, 
which we denote $\overline M^{\mu\nu}$, through
\ben
\overline M^{\mu\nu} = 
N!\de^{\mu\mu_2\ldots\mu_N\nu\nu_2\ldots\nu_N}
M_{\mu_2\nu_2} \ldots M_{\mu_N\nu_N}/\detbar M
\een
Thus we may write
\ben
\vev{\det h\,\pa_\mu\pa_\nu u^C} 
= \detbar M \left( \pa_\mu\pa_\nu u^C 
+ \frac{2}{N} \overline M^{\rho\si}\ga_{\rh\mu\nu}\pa_\si u^C
\right)
\een
and
\ben
\vev{\det h\,\pa_\ta u^C} 
= \detbar M \left( \pa_\mu\pa_\nu u^C 
+ \frac{2}{N} \overline M^{\rho\si}M_{\rh\ta}\pa_\si u^C
\right)
\een

\section{The mean field zero curvature equation}
Putting the results of Appendix (\ref{a:AveExtSurEqn}) together, we find that the 
Gaussian averaged equations for zero extrinsic curvature 
surfaces is
\bea
\left(g^{\mu\nu} - 
\frac{\pa}{\pa g_{\mu\nu}}
\right)
\detbar M \left[
\pa_\mu\pa_\nu u^C\right. &+& \frac{2}{N} \overline M^{\ka\ta} 
\ga_{\ka\mu\nu} \pa_\ta u^C \nonumber \\
- \Ga^\ta_{\mu\nu}\left(\pa_\ta u^C\right. &+& 
\left.\left.\frac{2}{N} \overline M^{\ka\la}M_{\la\ta}\pa_\la u^C \right)
\right] = 0
\label{e:AvSurf2}
\eea
For future convenience we will break this equation down into four terms:
\bea
T_A&=& \left( g^{\mu\nu} - 
\frac{\pa}{\pa g_{\mu\nu}}
\right)
\detbar M \pa_\mu\pa_\nu u^C\\
T_B&=& \frac{2}{N}\left( g^{\mu\nu} - 
\frac{\pa}{\pa g_{\mu\nu}}
\right)
\detbar M \overline M^{\ka\la} \ga_{\ka\mu\nu}\pa_\la u^C\\
T_C&=& \left( g^{\mu\nu} - 
\frac{\pa}{\pa g_{\mu\nu}}
\right)
\detbar M \Ga^\la_{\mu\nu}\pa_\la u^C\\
T_D&=& \frac{2}{N}\left( g^{\mu\nu} - 
\frac{\pa}{\pa g_{\mu\nu}}
\right)
\detbar M \overline M^{\ka\ta} M_{\ka\la} \Ga^\la_{\mu\nu}\pa_\ta u^C
\eea
Before reducing this equation further in cases of definite $N$  
we will need the following explicit expressions for the correlation 
functions $M_{\mu\nu}$ and $\ga_{\ka\mu\nu}$, consistent with 
the spatial O(D) symmetry:
\bea
M_{\mu\nu} & = & 
\left( 
  \ba{cc}
    T & 0 \\
    0 & S \de_{mn}
  \ea 
\right)\label{e:expM}\\
\ga_{000}  =  \half \dot T, \quad \ga_{0mn} &=& -\half \dot S\de_{mn},\quad
\ga_{m0n}  = \ga_{mn0} =  \half\dot S \de_{mn}
\label{e:expG}
\eea
Note that 
\ben
\gamma_{\mu\nu\rho} =
\half(M_{\mu\nu,\rho}+M_{\mu\rho,\nu}-M_{\nu\rho,\mu} ).
\een
We also need the Christoffel symbol for a flat FRW background,
which has metric $g_{\mu\nu} = a^2(\eta)\eta_{\mu\nu}$, where 
$\eta_{\mu\nu} = {\rm diag}(-1,1,1,1)$ is the Minkowski metric. It is 
\ben
\Ga^\la_{\mu\nu} = 
\frac{\dot a}{a}\left(
\de^\la_\mu\de^0_\nu + \de^\la_\nu\de^0_\mu - \eta_{\mu\nu} \eta^{\la 0}
\right)
\label{e:ChrisSymb}
\een

\section{Zero curvature for $N=1$}
The simplest case is with one coordinate field $u$, which is appropriate for
domain walls in 3 dimensions.  Here we have 
\bea
\detbar M &=& g^{\rh\si}M_{\rh\si} = M \label{e:WdetbarM}\\
\detbar M\overline M^{\ka\la} &=& g^{\ka\la} \label{e:WdetbarMM}\\
\detbar M \overline M^{\ka\la} M_{\ka\ta} &=& {M^\la}_\ta
\label{e:WdetbarMMM}
\eea
The required derivatives with respect to the metric are also easily
found:
\bea
\frac{\pa}{\pa g_{\mu\nu}}
\detbar M &=& M^{\mu\nu}
\label{e:Wterm1}\\
\frac{\pa}{\pa g_{\mu\nu}}
\detbar M \overline M^{\ka\la} &=& g^{\mu\ka}g^{\nu\la}
\label{e:Wterm2}
\eea
Using the explicit form of $M_{\mu\nu}$ 
(\ref{e:expM}) we can also write down
\ben
M = a^{-2}(-T + DS) \label{e:M}
\een

\subsubsection{Term $T_A$}
Using (\ref{e:WdetbarM}) and (\ref{e:Wterm1}) we find
\ben
T_A = (g^{\mu\nu}M - M^{\mu\nu})\pa_\mu\pa_\nu u.
\een
Using (\ref{e:expM}) and (\ref{e:M}) this simplifies to
\ben
T_A = -a^{-4}DS\left[ \ddot u - \left( \frac{D-1}{D} - 
\frac{T}{DS} \right) \nabla^2 u \right].
\een

\subsubsection{Term $T_B$}
Using (\ref{e:WdetbarMM}) and (\ref{e:Wterm1})  we have that
\ben
T_B = 2(g^{\mu\nu}g^{\ka\la} - g^{\mu\ka}g^{\nu\la})
\ga_{\ka\mu\nu}\pa_\la u.
\een
Using (\ref{e:expG}) one can quickly show that
\ben
T_B = 2a^{-4}D\dot S \dot u.
\een

\subsubsection{Term $T_C$}
Using (\ref{e:WdetbarM}), (\ref{e:Wterm1}) and (\ref{e:ChrisSymb}) one
finds
\bea
T_C &=& (g^{\mu\nu}M - M^{\mu\nu})\Ga^\la_{\mu\nu}\pa_\la u \nonumber\\
&=& [M(1-D)g^{0\la} - 2M^{0\la} + M g^{0\la}]\pa_\la u \nonumber \\
&=& a^{-4}DS[D-2 - (T/S)] \dot u.
\eea

\subsubsection{Term $T_D$}
Using (\ref{e:WdetbarMMM}) and (\ref{e:Wterm2}) we find that 
\bea
T_D &=& 2(g^{\mu\nu}g^{\ka\ta} - g^{\mu\ka}g^{\nu\ta})
M_{\ka\la}\Ga^\la_{\mu\nu}\pa_\ta u \nonumber\\
&=&2[(2-D)g^{0\la}g^{\ka\ta} - g^{\ka\la}g^{0\ta} - g^{\ta\la}g^{0\ka}]
M_{\ka\la}\left( \frac{\dot a}{a} \right)\pa_\ta u \nonumber \\
&=& 2[(1-D)M^{0\ta} - g^{0\ta}M]\left( \frac{\dot a}{a} \right)\pa_\ta
u.
\eea
Substituting the known forms of $M^{0\ta}$ and $M$ we arrive at
\ben
T_D = 2a^{-4}DS [1-(T/S)] \dot u.
\een

\subsection{The equation for $u$}
\label{a:ueqN1}
We can now construct the Gaussian averaged or ``mean field''
equations of motion satisfied by the coordinate function
$u$, which is applicable to domain walls when $D=3$. 
The equations are made from the 
four terms we calculated in the previous section:
$T_A+T_B-T_C-T_D=0$. Putting them all together, and 
dividing by the factor $a^{-4}DS$, we 
find
\ben
\ddot u^c + \frac{\mu}{\eta}\dot u^C - v^2 \nabla^2 u^C =0,
\een
with
\bea
\mu &=& [D - 3(T/S)]\left(\eta \frac{\dot a}{a} \right)
- 2 \left(\eta \frac{\dot S}{S} \right),\\
v^2 &=& [(D-1) - (T/S)]/D.
\eea
We can recover the well-known Allen-Cahn equation for 
the overdamped motion of domain walls by 
identifying the damping constant $\Ga = a/\dot a $, 
and neglecting $T/S$, $\eta \dot S/S$, and the second order 
time derivative of $u$:
\ben
\dot u = \Ga \frac{D-1}{D^2} \nabla^2 u.
\een

\section{Zero curvature for $N=2$}
When $N=2$ the expressions for the various quantities 
involving $M$ in the equations of motion are still
straightforward to evaluate:
\bea
\detbar M &=& (g^{\mu_1\nu_1}g^{\mu_2\nu_2} -
g^{\mu_1\nu_2}g^{\mu_2\nu_1}) M_{\mu_1\nu_1}M_{\mu_2\nu_2}\nonumber\\
&=& (M^2 - M^{\mu\nu}M_{\mu\nu})\label{e:SdetbarM}\\
\detbar M\overline M^{\ka\la} &=& (g^{\ka\la}g^{\mu_2\nu_2} -
g^{\ka\nu_2}g^{\mu_2\la})M_{\mu_2\nu_2} \nonumber\\
&=& g^{\ka\la}M - M^{\ka\la} \label{e:SdetbarMM}\\
\detbar M \overline M^{\ka\la} M_{\ka\ta} &=& (g^{\ka\la}g^{\mu_2\nu_2} -
g^{\ka\nu_2}g^{\mu_2\la})M_{\mu_2\nu_2}M_{\ka\ta} \nonumber\\
&=& M^\la_\ta M - M^{\ka\la}M_{\ka\ta},\label{e:SdetbarMMM}
\eea
We also need to differentiate two of these expressions with respect to the
metric $g_{\mu\nu}$.
\bea
\frac{\pa}{\pa g_{\mu\nu}}
\detbar M &=&
2(M^{\mu\nu}M-M^{\mu\la}{M_{\la}}^\nu)
\label{e:Sterm1}\\
\frac{\pa}{\pa g_{\mu\nu}}
\detbar M \overline M^{\ka\la} 
&=&
(g^{\mu\ka}g^{\nu\la}M + g^{\ka\la}M^{\mu\nu} - 
g^{\mu\ka}M^{\nu\la}-g^{\nu\la}M^{\mu\ka}).
\label{e:term2}
\eea
Introducing a further piece of 
notation, that $M\cdot M = {M^\mu}_\nu {M^{\nu}}_\mu$, we can 
show that
\bea
M\cdot M &=& a^{-4}(T^2 + D S^2) \label{e:MdotM}\\
\detbar M = M^2 - M\cdot M & = & a^{-4}DS^2[(D-1) - 2{T}/{S})] \label{e:detM}\\
M^{\mu\nu}M - M^{\mu\la}{M_\la}^\nu &=& 
-a^{-6}S^2\left(
\ba{cc}\!\!\!
D{T}/{S} & 0 \\
0 & \!\!\!\!\!\![(D-1) - (T/S)]\de_{mn}\!\!
\ea
\right) \label{e:MM}\\
g^{\mu\nu}\ga_{\ka\mu\nu} &=&
\half a^{-2}\de^0_\ka (\dot T + D \dot S) \label{e:ggamma}
\eea

\subsubsection{Term $T_A$}
Using \ref{e:Sterm1}, we find
\ben
T_A = (M^2 - M\cdot M) \pa^2 u^C 
- 2(M^{\mu\nu}M - M^{\mu\la}{M_\la}^\nu)\pa_\mu\pa_\nu u^C.
\een
Hence, using (\ref{e:detM}) and {\ref{e:MM}),
\ben
T_A = a^{-6} D(D-1)S^2
\left[
\ddot u^C - \left( \frac{D-2}{D} - \frac{2}{D}\frac{T}{S}
\right)\nabla^2 u^C.
\right]
\een

\subsubsection{Term $T_B$}
Using (\ref{e:SdetbarMM}) and (\ref{e:ggamma}) we find 
firstly that 
\ben
g^{\mu\nu} \detbar M \overline M^{\ka\la} \ga_{\ka\mu\nu} 
\pa_\la u^C = -\half a^{-6} DS(\dot T + D\dot S) \dot u^C. 
\een
Using (\ref{e:term2}) and (\ref{e:ggamma}) we find
\ben
\frac{\pa}{\pa g_{\mu\nu}}
\detbar M \overline M^{\ka\la} \ga_{\ka\mu\nu}\pa_\la u^C
= \half a^{-6} DS [ (D-2) \dot S - \dot T]\dot u^C.
\een
Putting the two expressions together we find
\ben
T_B = - a^{-6} D(D-1) S^2 \left(\frac{\dot S}{S} \right) \dot u^C
\een

\subsubsection{Term $T_C$}
From (\ref{e:SdetbarM}) and (\ref{e:Sterm1}) we can 
immediately write down 
\ben
T_C = \left( g^{\mu\nu}(M^2 - M\cdot M) - 
2M^{\mu\nu}M + 2M^{\mu\ka}{M_\ka}^\nu
\right)\Ga^\la_{\mu\nu}\pa_\la u^C.
\een
Using (\ref{e:ChrisSymb}) we find
\ben
T_C =
a^{-2}[(1-D)(M^2 - M\cdot M) + 4M^{00}(M^{00}-M) + 
2(M^2-M\cdot M)]\left(\frac{\dot a}{a}\right) \dot u^C.
\een
With equations (\ref{e:SdetbarM}) and (\ref{e:MM})
we arrive at
\ben
T_C = -a^{-6}D(D-1)S^2[(D-3) - 2(T/S)]
\left(\frac{\dot a}{a}\right)\dot u^C.
\een

\subsubsection{Term $T_D$}
For this last term we begin with
\ben
g^{\mu\nu} \Ga^\la_{\mu\nu} = (1-D)
\left(\frac{\dot a}{a}\right) g^{0\la}.
\een
Hence from (\ref{e:SdetbarMMM}) we see that 
\bea
g^{\mu\nu}\detbar M \overline M^{\ka\ta}{M_{\ka\la}}
\Ga^\la_{\mu\nu}\pa_\la u^C &= &
a^{-2}(g^{00}M-M^{00}){M_0}^0(1-D)
\left(\frac{\dot a}{a}\right)\dot u^C \nonumber\\
&=& a^{-6}D(D-1)ST\left(\frac{\dot a}{a}\right)\dot u^C.
\label{e:termD1}
\eea
The second term in expression D is more complicated. 
From (\ref{e:term2}) and (\ref{e:ChrisSymb}) we have
\bea
g^{\mu\rh}g^{\nu\si} \frac{\pa}{\pa g^{\rh\si}}
\detbar M \overline M^{\ka\ta}M_{\ka\la} \Ga^\la_{\mu\nu}
&=& 
(g^{\mu\ka}g^{\nu\ta} M + g^{\ka\ta}M^{\mu\nu} \nonumber\\
&-& g^{\mu\ka}M^{\nu\ta} - g^{\nu\ta}M^{\mu\ka})
M_{\ka\la}\Ga^\la_{\mu\nu}.
\eea
After some algebra we find
\bea
\detbar M \overline M^{\ka\ta}M_{\ka\la} \Ga^\la_{\mu\nu} 
&=& [g^{0\ta}(M^2 - M\cdot M) - 2M^{0\la}M + 2 {M^\ta}_\la 
M^{0\la}] \left(\frac{\dot a}{a}\right) \nonumber \\
&=&a^{-6}D(D-1)S^2\de_0^\ta.
\label{e:termD2})
\eea
Subtracting (\ref{e:termD2}) multiplied by 
$\pa_\ta u^C$ from (\ref{e:termD2}), we
arrive at
\ben
T_D = -a^{-6}D(D-1)S^2[1-(T/S)] \left(\frac{\dot a}{a}\right)
\dot u^C.
\een

\subsection{The equation for $u^C$ ($N=2$)}
\label{a:ueqN2}
We can now construct the Gaussian averaged or ``mean field''
equations of motion satisfied by the coordinate functions 
$u^C$ in the case $N=2$, appropriatre for strings in 3 
spatial dimensions. The equations are made from the 
four terms we calculated in the previous section:
$T_A+T_B-T_C-T_D=0$. Putting them all together, and 
dividing by the common factor $a^{-6}D(D-1)S^2$, we 
find
\ben
\ddot u^c + \frac{\mu}{\eta}\dot u^C - v^2 \nabla^2 u^C =0,
\een
with
\bea
\mu &=& [D-2 - 3(T/S)]\left(\eta \frac{\dot a}{a} \right)
- \left(\eta \frac{\dot S}{S} \right),\\
v^2 &=& [(D-2) - 2(T/S)]/D.
\eea
We can recover the results of Toyoki and Honda for the motion of
overdamped strings in $D=3$ by setting their diffusion constant $\Gamma
= a/\dot a$, and neglecting $T/S$ and $\eta \dot S/S$. In this case we
get 
\ben
\dot u^C = \frac{\Ga}{3} \nabla^2 u^C,
\een
which is identical to their equation (3.10).

\section{Probability distribution for $F_{ij}$}
\label{a:ProDisF}
The definition of the antisymmetric tensor $F_{ij}$ is 
\ben
F_{ij} = \pa_iu^A\pa_ju^B\ep_{AB}.
\een
The probability ditribution for $F_{ij}$ is therefore constructed from the
Gaussian probability 
distribution of $\pa_iu^A$.
$F_{ij}$ is antisymmetric, so we need only consider half 
of the non-zero elements, e.g.~by imposing $i<j$.
Moreover, it is convenient to scale out the variance 
of $\pa_iu^A$, defining variables $\pi_i^A$ and $f_{ij}$ as 
follows:
\ben
\pa_iu^A = \sqrt{S} \pi_i^A, \quad F_{ij} = S f_{ij}
\een
where
\ben
\vev{\pa_iu^A(x)\pa_ju^B(x)} = S(t)\de_{ij}\de^{AB}.
\een
Hence, the probability distribution for $f_{ij}$ is
\ben
\left. P(f_{ij}) \right|_{i<j} = 
\int \prod_A \frac{d^D \pi_i^A}{(2\pi)^{\frac{D}{2}} }
e^{-\half \pi_i^A\pi_i^A} \de (f_{ij} - \pi_i^A\pi_j^B\ep_{AB})|_{i<j}.
\een
Using the Fourier representation of the $\de$-function,
\ben
\left. P(f_{ij}) \right|_{i<j} = \int \frac {d^P k}{(2\pi)^P}
\int \prod_A \frac{d^D \pi_i^A}{(2\pi)^{\frac{D}{2}}} 
e^{-\half \pi_i^A\pi_i^A + i \sum_{i<j} k^{ij}(f_{ij} - 
\pi_i^A\pi_j^B\ep_{AB}) },
\een
where $P=D(D-1)/2$ is the dimension of $k_{ij}$.

We now do the $\pi_i^A$ integrations in turn, starting with 
the highest $A$. First, note that 
\ben
\left. P(f_{ij}) \right|_{i<j} = \int \frac {d^P k}{(2\pi)^P}
e^{i \sum_{i<j} k^{ij}f_{ij}}
\int \prod_A \frac{d^D \pi_i^A}{(2\pi)^{\frac{D}{2}} }
e^{-\half \pi_i^A\pi_i^A - i k^{ij}
\pi_i^1\pi_j^2 },
\een
where there is now no restriction on the sum over $i,j$ 
in the second exponential. Second,
define the variable $q^j = k^{ij}\pi_i^1$.  Then we have to
evaluate the integral
\ben
I(k_{ij}) =  \int \prod_A \frac{d^D \pi_i^A}{(2\pi)^{\frac{D}{2}} }
e^{-\half \pi_i^A\pi_i^A -iq^j\pi_j^2}.
\een
Doing the $\pi_i^2$ integral first, this is
\bea
I(k_{ij}) &=&  \int \frac{d^D \pi_i^1}{(2\pi)^{\frac{D}{2}} }
e^{-\half \pi_i^1M_{ij}\pi_j^1}, \\
&=& {\det}^{-\half} M 
\eea
where 
\ben
M_{ij} = \de_{ij} + k_{ik}k_{jk}.
\een
At this point we specialise to 3D, where we can write
\ben
k_{ij} = \ep_{ijk}p_k.
\een
Hence
\ben
M_{ij} = \de_{ij}(1+p^2) - p_ip_j.
\een
The eigenvalues of this matrix  are $1+p^2$ (twice) and 
$1$, so 
\ben
{\det}^{-\half} M = (1+p^2)^{-1}.
\een
Thus the probability distribution of $f_{ij}$ is 
\ben
\left. P(f_{ij}) \right|_{i<j} = \int \frac {d^3 p}{(2\pi)^3} 
\frac{1}{1+p^2} e^{i\sum_{i<j}\ep^{ijk}f_{ij}p_k}.
\label{e:PF}
\een
In 3D we can replace $f_{ij}$ by $\phi_k = 
\sum_{i<j}\ep^{ijk}f_{ij}$, and the integral may be easily evaluated to give
\ben
P(\phi_k) = \frac{1}{4\pi\phi}e^{-\phi},
\een
where $\phi^2 = \phi_k\phi_k$. 

\section{Integral formulae and correlation functions}
\label{a:Integ}
In this appendix we perform the integrations necessary to evaluate the
functions $C$, $S$ and $T$, defined in Section, which we repeat here for
convenience.
\begin{eqnarray}
\delta^{AB}
C(\eta) &=& \langle u^A(x)u^B(x)\rangle,\nonumber\\
\delta^{AB} M_{\mu\nu}(\eta)&=&\langle\partial_\mu u^A(x)
\partial_\nu u^B(x)\rangle,\nonumber
\end{eqnarray}
with
\begin{equation}
M_{\mu\nu} = \left(\begin{array}{cc}
		T(\eta) & 0 \cr
		0 & \delta_{mn} S(\eta) 
             \end{array}\right).
\end{equation}
We shall also evaluate $\dot C$ for the mixed correlator 
$\vev{\pa_\mu u^A u^B}$.
We recall from Eq.\ (\ref{e:LinuEq})
that the linearised solution for $u^A$ with the correct boundary
conditions is 
\ben
u^C_{\bk}(\eta)  = A^C_{\bk}\left(\frac{\eta}{\eta_{\rm i}}\right)^{(1-
\mu)/2+\nu} \frac{J_\nu(kv\eta)}{(kv\eta)^\nu},
\een
with $\nu = \pm (1-\mu)/2$.  If we demand regularity and
convergent integrals as $\eta \to 0$, 
we must take the negative sign here, as it will turn out that $\mu > 1$.

In order to calculate the two-point functions it is useful to define 
the following integral
\ben
I(\rho,\si,\ta) \equiv 
\int_0^\infty dz\, z^{-\rho} J_\si(z) J_\ta(z),
\een
which has the value  \cite{GraRyz}
\ben
I(\rho,\si,\ta) = \frac{1}{2^\rho}
\frac{
\Ga(\rh)\Ga\left(\frac{\si+\ta-\rh+1}{2}\right)}
{
\Ga\left(\frac{\rh-\si+\ta+1}{2}\right)
\Ga\left(\frac{\rh+\si+\ta+1}{2}\right)
\Ga\left(\frac{\rh+\si-\ta+1}{2}\right)
},
\een
provided ${\rm Re}(\si+\ta+1) >
{\rm Re}(\rh) > 0$.  The first inequality comes from the condition that the
integral be defined as $z \to 0$, and the second from requiring that it
converge as $z\to\infty$. There is a
simple pole in at $\rho = 0$. We can see that this comes from
the $z^{-1/2}$ behaviour of the Bessel functions as $z\to\infty$,
and  corresponds
to a
logarithmically divergent integral.

Defining the Fourier transform of the correlator $C$ in the usual way through
\ben
C(\eta) = \int\frac{d^D k}{(2\pi)^D} C_{\bk}(\eta),
\een
we see from the solutions for $u^A$ that
\ben
C(\eta) = 
\frac{1}{(v\eta)^D}\frac{\Om_D}{(2\pi)^D}
\int dz \, z^{D-1-2\nu}J^2_\nu(z) P_A({\bk})
\een
where $z = kv\eta$, and $\Om_D = 2\pi^{D/2}/\Ga(D/2)$ is the volume element of
a $(D-1)$-sphere. We assume a power-law form for the power spectrum 
of $A_{\bk}^A$,
\ben
P_A({\bk}) = \frac{\si_{\rm i}(2\pi)^D}{\Om_D\La^D\Ga(D+q)}
\left(\frac{k}{\La}\right)^q e^{-k/\La},
\een
where $\La$ is a high wavenumber cut-off, satisfying $\La v \eta \gg 1$ for all
$\eta$ of interest, and $\si_{\rm i}$ is the variance.
 Hence, defining $\be = 2\nu -D - 1 -q$, 
\ben
C(\eta) = 
\frac{v^q}{(\La v\eta)^{D+q}}\frac{\si_{\rm i}}{\Ga(D+q)}
I(2+\be,\nu,\nu).
\een
Let us now calculate $S$ from
\ben
DS(\eta) = \int \frac{d^Dk}{(2\pi)^D} k^2C_{\bk}(\eta).
\een
One can straightforwardly show that
\ben
DS = 
\frac{v^q}{(\La v\eta)^{D+q}}\frac{1}{(v\eta)^2} 
\frac{\si_{\rm i}}{\Ga(D+q)}
I(\be,\nu,\nu).
\een
The correlation function $T$ is obtained from
\ben
\de^{AB} T = 
\int \frac{d^Dk}{(2\pi)^D} \vev{\dot u^A_{\bk}(\eta)
\dot u^B_{-\bk}(\eta)}.
\een
Given the identity \cite{GraRyz}
\ben
\frac{d}{dz}\left(\frac{J_\nu(z)}{z^\nu}\right) = - \frac{J_{\nu+1}(z)}{z^\nu},
\een
one can show that
\ben
T = 
\frac{v^q}{(\La v\eta)^{D+q}}\frac{1}{\eta^2} 
\frac{\si_{\rm i}}{\Ga(D+q)}
I(\be,\nu+1,\nu+1).
\een
Note that the ratios $S/C$ and $T/S$ depend on the initial conditions only
though the power $q$, which appears in $\be$;
\bea
\frac{S}{C} &=& \frac{1}{D(v\eta)^2}\frac{I(\be,\nu,\nu)}{I(2+\be,\nu,\nu)}\\
\frac{T}{S} &=& {Dv^2}\frac{I(\be,\nu+1,\nu+1)}{I(\be,\nu,\nu)}.
\eea
A little more algebra shows that
\bea
\frac{S}{C}  &=&  \frac{1}{\eta^2} \frac{D+2+\beta}{4v^2}\frac{\be+1}{\be},\\
\frac{T}{S}  &=&  Dv^2\frac{D+2}{D+2+2\be}.
\eea
Note that the ratio $S/C$ appears to have a simple pole at $\be=0$: however,
when the cut-off is in place this is replaced by a logarithm, with 
\ben
\frac{S}{C} \sim \frac{1}{\eta^2} \log (\La v \eta).
\een


\begin{thebibliography}{99}

\bibitem{BarAsy}
V.~A.~Rubakov and M.~E.~Shaposhnikov,
Usp.\ Fiz.\ Nauk {\bf 166}, 493 (1996)
[Phys.\ Usp.\  {\bf 39}, 461 (1996)]
[arXiv:hep-ph/9603208];
A.~Riotto and M.~Trodden,
Ann.\ Rev.\ Nucl.\ Part.\ Sci.\  {\bf 49}, 35 (1999)
[arXiv:hep-ph/9901362].

\bibitem{DenFlu}
R.~Durrer, M.~Kunz and A.~Melchiorri,
Phys.\ Rept.\  {\bf 364} (2002) 1
[arXiv:astro-ph/0110348].

\bibitem{MagFie}
D.~Grasso and H.~R.~Rubinstein,
Phys.\ Rept.\  {\bf 348}, 163 (2001)
[arXiv:astro-ph/0009061].

\bibitem{Boyanovsky:1999db}
D.~Boyanovsky and H.~J.~de Vega,
arXiv:hep-ph/9909372.

\bibitem{HotAbHiggs}
A.~Rajantie and M.~Hindmarsh,
Phys.\ Rev.\ D {\bf 60}, 096001 (1999)
[arXiv:hep-ph/9904270];
%
M.~Hindmarsh and A.~Rajantie,
Phys.\ Rev.\ Lett.\  {\bf 85}, 4660 (2000)
[arXiv:cond-mat/0007361];
M.~Hindmarsh and A.~Rajantie,
Phys.\ Rev.\ D {\bf 64}, 065016 (2001)
[arXiv:hep-ph/0103311].

\bibitem{Rajantie:2001ps}
A.~Rajantie,
Int.\ J.\ Mod.\ Phys.\ A {\bf 17}, 1 (2002)
[arXiv:hep-ph/0108159].

\bibitem{BraGas} 
R.~H.~Brandenberger and C.~Vafa,
Nucl.\ Phys.\ B {\bf 316}, 391 (1989);
S.~Alexander, R.~H.~Brandenberger and D.~Easson,
Phys.\ Rev.\ D {\bf 62}, 103509 (2000)
[arXiv:hep-th/0005212].

\bibitem{Bra94}
A.J.~Bray, Adv.\ Phys.\  {\bf 43}, 357 (1994).

\bibitem{AllCah79} 
S.M.~Allen and J.W.~Cahn,
Acta.\ Metal.\ {\bf 27} 1085 (1979).

\bibitem{MotMeaCur}
L.~Bronsard and R.V.~Kohn,
J.\ Diff.\ Eq.\ {\bf 90} 211 (1991);
L.C.~Evans and J.~Spruck, 
J.\ Diff.\ Geom.\ {\bf 33} 635 (1991);
L.C.~Evans, H.M.~Soner and P.E.~Souganis,
Comm. Pure Appl. Math. {\bf 45} 1097 (1992).

\bibitem{VilShe94} A. Vilenkin and E.P.S. Shellard, {``Cosmic Strings 
        and Other Topological Defects''} (Cambridge Univ. Press, Cambridge,
        1994) 

\bibitem{Hindmarsh:1994re}
M.~B.~Hindmarsh and T.~W.~Kibble,
Rept.\ Prog.\ Phys.\  {\bf 58}, 477 (1995)
[arXiv:hep-ph/9411342].

\bibitem{KolTur90}{E.W. Kolb and M.S. Turner},  
        {``The Early  Universe''} {(Addison-Wesley, Redwood City, 1990)} 

\bibitem{Turok:1991qq}
N.~Turok and D.~N.~Spergel,
Phys.\ Rev.\ Lett.\  {\bf 66}, 3093 (1991).

\bibitem{FilBra94} J.A.N. Filipe and A.J. Bray, {\it Phys. Rev.}  
        {\bf E50}, {2523} (1994) [hep-ph/9605346]

\bibitem{Kunz:1996ka}
M.~Kunz and R.~Durrer,
Phys.\ Rev.\ D {\bf 55}, 4516 (1997)
[arXiv:astro-ph/9612202].

\bibitem{Boyanovsky:1998yp}
D.~Boyanovsky, H.~J.~de Vega, R.~Holman and J.~Salgado,
Phys.\ Rev.\ D {\bf 59}, 125009 (1999)
[arXiv:hep-ph/9811273].

\bibitem{Bennett:1993fy}
D.~P.~Bennett and S.~H.~Rhie,
structure formation,''
Astrophys.\ J.\  {\bf 406}, L7 (1993) [hep-ph/9207244].

\bibitem{Durrer:1994da}
R.~Durrer, A.~Howard and Z.~Zhou,
Phys.\ Rev.\  {\bf D49}, 681 (1994) [astro-ph/9311040].

\bibitem{Pen:1994nx}
U.~Pen, D.~N.~Spergel and N.~Turok,
ordering,''
Phys.\ Rev.\  {\bf D49}, 692 (1994).

\bibitem{Press:1989yh}
W.~H.~Press, B.~S.~Ryden and D.~N.~Spergel,
Astrophys.\ J.\ {\bf 347}, 590 (1990).

\bibitem{Coulson:1996nv}
D.~Coulson, Z.~Lalak and B.~Ovrut,
Phys.\ Rev.\  {\bf D53}, 4237 (1996).

\bibitem{Larsson:1997sp}
S.~E.~Larsson, S.~Sarkar and P.~L.~White,
Phys.\ Rev.\  {\bf D55}, 5129 (1997) [hep-ph/9608319].

\bibitem{Casini:2001ai}
H.~Casini and S.~Sarkar,
Phys.\ Rev.\ D {\bf 65}, 025002 (2002)
[arXiv:hep-ph/0106272].

\bibitem{Vincent:1998cx}
G.~Vincent, N.~D.~Antunes and M.~Hindmarsh,
Phys.\ Rev.\ Lett.\  {\bf 80}, 2277 (1998) [hep-ph/9708427].


\bibitem{Moore:2001px}
J.~N.~Moore, E.~P.~Shellard and C.~J.~Martins,
Phys.\ Rev.\ D {\bf 65}, 023503 (2002)
[arXiv:hep-ph/0107171].

\bibitem{Ryden:1989vj}
B.~S.~Ryden, W.~H.~Press and D.~N.~Spergel,
Astrophys. J. {\bf 357}, 293 (1990).

\bibitem{Yamaguchi:2000dy}
M.~Yamaguchi, J.~Yokoyama and M.~Kawasaki,
Phys.\ Rev.\  {\bf D61}, 061301 (2000) [hep-ph/9910352].


\bibitem{GarHin02} T. Garagounis and M. Hindmarsh, in preparation (2002).

\bibitem{Carter:1994ag}
B.~Carter and R.~Gregory,
Phys.\ Rev.\ D {\bf 51}, 5839 (1995)
[arXiv:hep-th/9410095].

\bibitem{Anderson:1997ip}
M.~Anderson, F.~Bonjour, R.~Gregory and J.~Stewart,
Phys.\ Rev.\ D {\bf 56}, 8014 (1997)
[arXiv:hep-ph/9707324].

\bibitem{Arodz:1997va}
H.~Arodz,
Nucl.\ Phys.\ B {\bf 509}, 273 (1998)
[arXiv:hep-th/9703168].

\bibitem{Albrecht:1989mk}
A.~Albrecht and N.~Turok,
Phys.\ Rev.\ D {\bf 40}, 973 (1989).

\bibitem{StrSimBB} 
D.~P.~Bennett and F.~R.~Bouchet,
Phys.\ Rev.\ Lett.\  {\bf 63}, 2776 (1989);
D.~P.~Bennett and F.~R.~Bouchet,
Phys.\ Rev.\ D {\bf 41}, 2408 (1990);
D. Bennett, in  
        ``The Formation and Evolution of Cosmic Strings'' eds.  
        G. Gibbons, S.W. Hawking, and T. Vachaspati,  
        (Cambridge Univ. Press, Cambridge, 1990); F. Bouchet, {\it ibid.}.  
        
\bibitem{StrSimAS}
B.~Allen and E.~P.~Shellard,
Phys.\ Rev.\ Lett.\  {\bf 64}, 119 (1990);
E.~P.~Shellard and B.~Allen, 
        in  
        ``The Formation and Evolution of Cosmic Strings'' eds.  
        G. Gibbons, S.W. Hawking, and T. Vachaspati,  
        (Cambridge Univ. Press, Cambridge, 1990)

\bibitem{Vincent:1996rb}
G.~R.~Vincent, M.~Hindmarsh and M.~Sakellariadou,
Phys.\ Rev.\ D {\bf 56}, 637 (1997)
[arXiv:astro-ph/9612135].

\bibitem{Kibble:1984hp}
T.~W.~Kibble,
Nucl.\ Phys.\ B {\bf 252}, 227 (1985)
[Erratum-ibid.\ B {\bf 261}, 750 (1985)].

\bibitem{KibCop}{T.W.B. Kibble and E. Copeland 1991}, {\it Physica 
        Scripta} {\bf T36}, {153} (1991); {E. Copeland, T.W.B. Kibble  
        and D. Austin}, {\it Phys. Rev.} {\bf D45}, {R1000} (1992);  
        {D. Austin, E. Copeland and T.W.B. Kibble}, {\it Phys. Rev.} 
        {\bf D48}, {5594} (1993). 

\bibitem{VelDep1Scale} 
C.~J.~Martins and E.~P.~Shellard,
Phys.\ Rev.\ D {\bf 54}, 2535 (1996)
[arXiv:hep-ph/9602271];
C.~J.~Martins and E.~P.~Shellard,
Phys.\ Rev.\ D {\bf 65}, 043514 (2002)
[arXiv:hep-ph/0003298].

\bibitem{Hindmarsh:1996xv}
M.~Hindmarsh,
Phys.\ Rev.\ Lett.\  {\bf 77}, 4495 (1996)
[arXiv:hep-ph/9605332].

\bibitem{HinCop97}  M.H.\ and E.J. Copeland
	``Defects without Computers''
	in `Topological defects in Cosmology', eds F. Melchiorri and M. Signore, 
	(World Scientific, Singapore, 1997).

\bibitem{OhtJasKaw82} T. Ohta, D. Jasnow, and K. Kawasaki,  
        {Phys. Rev. Lett.} {\bf  49}, 1223 (1982). 

\bibitem{DefScale} A. Bray, in  
        ``Formation and Interactions of Topological Defects'',  
        NATO ASI Series, eds. R. Brandenberger and  
        A-C. Davis, (Plenum, New York, 1995); N. Goldenfeld, {\it ibid.};  
        G. Mazenko, {\it ibid.} 


\bibitem{Olum:1999sg}
K.~D.~Olum and J.~J.~Blanco-Pillado,
Phys.\ Rev.\ Lett.\  {\bf 84}, 4288 (2000)
[arXiv:astro-ph/9910354].

\bibitem{Carter:2000wv}
B.~Carter,
Int.\ J.\ Theor.\ Phys.\  {\bf 40}, 2099 (2001)
[arXiv:gr-qc/0012036].

\bibitem{Moss:1998jf}
I.~G.~Moss and N.~Shiiki, 
Nucl.\ Phys.\ B {\bf 565}, 345 (2000)
[arXiv:hep-th/9904023].

\bibitem{Coulson:1995nv}
D.~Coulson, Z.~Lalak and B.~A.~Ovrut,
Phys.\ Rev.\ D {\bf 53}, 4237 (1996).

\bibitem{Larsson:1996sp}
S.~E.~Larsson, S.~Sarkar and P.~L.~White,
Phys.\ Rev.\ D {\bf 55}, 5129 (1997)
[arXiv:hep-ph/9608319].

\bibitem{ToyHon87} H. Toyoki and K. Honda, {\it Prog. Theor. Phys.}  
        {\bf 78}, 237 (1987). 

\bibitem{LiuMaz92} F. Liu and G.F. Mazenko, {\it Phys. Rev.}  
        {\bf B46}, 5963 (1992). 

\bibitem{Perc} 
T.~Vachaspati,
Phys.\ Rev.\ D {\bf 44}, 3723 (1991);
        R.M. Bradley, P.N. Strenski, and J-M. Debierre,  
        {Phys.\ Rev.\ } {\bf A45}, {8513} (1992);  
M.~Hindmarsh and K.~Strobl,
Nucl.\ Phys.\ B {\bf 437}, 471 (1995)
[arXiv:hep-th/9410094].

\bibitem{ToyHon86} H. Toyoki and K. Honda, {\it Phys Rev.} {\bf B33},  
        385 (1986). 

\bibitem{MonGol92} M. Mondello and N. Goldenfeld, {\it Phys. Rev.} {\bf A45},  
        657 (1992). 

\bibitem{FilBraPur95} J.A.N. Filipe, A.J. Bray, and S. Puri, {\it Phys. 
        Rev.} {\bf E52}, 6082 (1995).
        [arXiv:cond-mat/9504079] 

\bibitem{Scherrer:1997sq}
R.~J.~Scherrer and A.~Vilenkin,
Phys.\ Rev.\ D {\bf 58}, 103501 (1998)
[arXiv:hep-ph/9709498].

\bibitem{GraRyz} I.S. Gradshteyn and I.M. Ryzhik, ``Table of series, integrals
and products, 5th edition'' (Academic Press, San Diego, 1996)


\bibitem{Vac90} T. Vachaspati, Phys. Lett. B242, 41 (1990).

\bibitem{Arodz:1995pt}
H.~Arodz and L.~Hadasz,
Phys.\ Rev.\ D {\bf 54}, 4004 (1996)
[arXiv:hep-th/9506021].

\end{thebibliography}
\end{document}